\documentclass[11pt]{article}

\usepackage[dvips]{color}

\definecolor{kernel}{rgb}{0.5,0.5,0}
\newcommand{\markkern}{\scriptstyle \color{kernel}}

\newcommand{\bm}[1]{\mbox{\boldmath $#1$}}

\def\rr{R}
\def\sig{f}
\def\dersigo{{\sig'_c}}
\def\sigo{\sig_c}
\def\dersig{\sig'}
\def\curv{\epsilon}              

\def\difcqcqin{\delta Q}

\def\ttauout{T^+}
\def\cqout{Q^+}
\def\tone{{\cal T}^{+}_1}
\def\ttwo{{\cal T}^{+}_2}

\def\Fqout{F^q_{+}}
\def\Gqout{G^q_{+}}
\def\Pqout{P^q_{+}}
\def\Rqout{R^q_{+}}
\def\Hqout{H^q_{+}}

\def\Fextout{F^{\kappa}_{+}}
\def\Gextout{G^{\kappa}_{+}}
\def\Pextout{P^{\kappa}_{+}}
\def\Rextout{R^{\kappa}_{+}}
\def\Hextout{H^{\kappa}_{+}}

\def\cy{Y}
\def\ppi{\Pi}
\def\cchi{\chi}
\def\ch{{\cal H}}

\def\yone{{{\cal Y}_1}}
\def\ytwo{{{\cal Y}_2}}
\def\wone{{{\cal W}_1}}
\def\wzero{W}
\def\wtwo{{{\cal W}_2}}
\def\qone{{{\cal Q}_1}}
\def\qtwo{{{\cal Q}_2}}
\def\uone{{{\cal U}_1}}
\def\utwo{{{\cal U}_2}}
\def\bone{{{\cal B}_1}}
\def\btwo{{{\cal B}_2}}
\def\wchi{{\tilde{W}_{R}}}
\def\ychi{{Y_{R}}}

\def\asig{a_{\Sigma}}

\def\derxichi{\cchi'}
\def\derxixichi{\cchi''}
\def\derxixixichi{\cchi'''}

\def\derxiyone{\yone'}
\def\derxiytwo{\ytwo'}
\def\derxiwone{\wone'}
\def\derxiwzero{\wzero'}

\def\derxiwtwo{\wtwo'}
\def\derxiqone{\qone'}
\def\derxiqtwo{\qtwo'}
\def\derxiuone{\uone'}
\def\derxiutwo{\utwo'}
\def\derxiwchi{\wchi'}  
\def\derxiychi{\ychi'}

\def\derxichi{\cchi'}
\def\derxixichi{\cchi''}
\def\derxixixichi{\cchi'''}
\def\derxipsi{\Psi'}
\def\derxiphi{\Phi'}
\def\derxich{\ch'}

\def\cqin{Q^-}
\def\ttauin{T^-}

\def\tinone{{\cal T}^{-}_1}
\def\tintwo{{\cal T}^{-}_2}

\def\uperonepsup{{U^{(1)}}}
\def\deruperonepsupr{\frac{\partial \uperonepsup}{\partial r} }
\def\cpsup{{\cal P}}
\def\cgsup{{\cal G}}
\def\dercpsupr{ \frac{\partial \cpsup}{\partial r} }
\def\der2cpsupr{ \frac{\partial^2 \cpsup}{\partial r^2} }
\def\dercgsupr{ \frac{\partial \cgsup}{\partial r} }

\def\Fqin{F^q_{-}}
\def\Gqin{G^q_{-}}
\def\Pqin{P^q_{-}}
\def\Rqin{R^q_{-}}
\def\Hqin{H^q_{-}}

\def\Fextin{F^{\kappa}_{-}}
\def\Gextin{G^{\kappa}_{-}}
\def\Pextin{P^{\kappa}_{-}}
\def\Rextin{R^{\kappa}_{-}}
\def\Hextin{H^{\kappa}_{-}}

\def\DeltaS{\Delta_{S^2}}
\def\asigdot{\dot\asig}
\def\asigdotdot{\ddot\asig}

\def\cs{\frac{\cos \theta}{\sin \theta}}

\def\K{\mathcal{K}}
\def\gback{g^{(0)}}

\def\gfpert{g^{(1)}}

\def\supo{\Sigma_0}

\def\domegas{(d\vartheta^2 + \sin^2 \vartheta d\varphi^2)}

\def\eqq{\stackrel{\supo}{=}}

\def\ddl{{\frac{d}{d\lambda}}}
\def\Fqo{{\markkern F^q_{0}}}
\def\Gqo{{\markkern G^q_{0}}}
\def\Pqo{{\markkern P^q_{0}}}
\def\Pqm{{ \markkern P^q_{m}}}
\def\Rqo{{ \markkern R^q_{0}}}
\def\Rqone{{ \markkern R^q_{1}}}
\def\Rqtwo{{ \markkern R^q_{2}}}
\def\Rqthree{{ \markkern R^q_{3}}}
\def\Rqm{{ \markkern R^q_{m}}}
\def\kernelPq{{\markkern \Pqo+\Pqm Y_1^m}}
\def\kernelRq{{\markkern \Rqo+\Rqm Y_1^m}}
\def\Fexto{{ \markkern F^{\kappa}_{0}}}
\def\Gexto{{ \markkern G^{\kappa}_{0}}}
\def\Pexto{{ \markkern P^{\kappa}_{0}}}
\def\Pextm{{ \markkern P^{\kappa}_{m}}}
\def\Rexto{{ \markkern R^{\kappa}_{0}}}
\def\Rextone{{ \markkern R^{\kappa}_{1}}}

\def\Rextm{{ \markkern R^{\kappa}_{m}}}

\def\kernelRext{{ \markkern \Rexto+\Rextm Y_1^m}}
\def\opthp{\triangle_{\theta}\cpsup}

\def\uytwo{\utwo + \ytwo - 
{\markkern (\kernelRq)}}
\def\wwtwo{\wtwo - \Gqo -\ddl[\uytwo|_{\supo}]}
\def\duytwo{\derxiutwo + \derxiytwo - {\markkern 2\asig} (\kernelRext)}
\def\dwwtwo{\derxiwtwo -{\markkern 2 \asig \Gexto} - \ddl[\duytwo|_{\supo}]}

\def\diffqq{\frac{\difcqcqin}{\asig}}

\def\uyone{\uone + \yone +\frac{1}{2} \cchi - {\markkern  (\kernelPq)}}
\def\wwone{\wone +\wzero  - \Fqo -\ddl[\uyone|_{\supo}]}
\def\duyone{\derxiuone + \derxiyone - {\markkern  2\asig (\kernelRext)}}
\def\dwwone{\derxiwone -\wchi -{\markkern 2 \asig \Fexto} - \ddl[\duyone|_{\supo}]
-\ddl\left(\diffqq\right)}

\def\uytwoP{\utwo - {\markkern  (\kernelRq)}}
\def\wwtwoP{\wtwo - \Gqo -\ddl[\uytwoP|_{\supo}]}
\def\duytwoP{\derxiutwo - {\markkern  2\asig (\kernelRext)}}
\def\dwwtwoP{\derxiwtwo - {\markkern   2\asig \Gexto}
 - \ddl[\duytwoP|_{\supo}]}

\def\uyoneP{\uone - {\markkern  (\kernelPq)}}
\def\wwoneP{\wone - \Fqo -\ddl[\uyoneP|_{\supo}]}
\def\duyoneP{\derxiuone - {\markkern  2\asig (\kernelRext)}}
\def\dwwoneP{\derxiwone -\wchi - {\markkern  2 \asig \Fexto} - \ddl[\duyoneP|_{\supo}]
-\ddl\left(\diffqq\right)}

\def\nback{n^{(0)}}

\def\vnback{\vec n^{(0)}}

\def\fnback{\bm n^{(0)}}
\def\fffback{{q^{(0)}}}
\def\ffffpert{q^{(1)}}
\def\sffback{{k^{(0)}}}
\def\sfffpert{k^{(1)}}

\def\Journal#1#2#3#4#5#6{``#6'', {#1} {\bf #2}, #3#4 (#5)}

\def\CQG{\em Class. Quantum Grav.}

\def\PRD{\em Phys. Rev. D }
\def\GRG{\em Gen. Rel. Grav.}

\def\IJMPD{\em Int. J. Mod. Phys. D}
\def\PR{\em Phys. Rev.}
\def\RMP{\em Rev. Mod. Phys.}

\def\JMP{\em J. Math. Phys.}

\def\PRL{\em Phys. Rev. Lett.}

\def\AP{\em Ap. J.}

\def\PHLL{\em Phys. Lett. A }

\newtheorem{theorem}{Theorem}[section]

\begin{document}

\title{First order perturbations 
of the Einstein-Straus and Oppenheimer-Snyder models}


\author{
Marc Mars\footnotemark[1], 
Filipe C. Mena\footnotemark[2] and
Ra\"ul Vera\footnotemark[3] \\
{\small \footnotemark[1] Facultad de Ciencias, Universidad de Salamanca, 
Plaza de
la Merced s/n, 37008 Salamanca, Spain} \\
{\small \footnotemark[2] Departamento de Matem\'atica, Universidade do Minho, 4710 Braga,
Portugal}
\\
{\small \footnotemark[3] Fisika Teorikoaren Saila, Euskal Herriko Unibertsitatea,
644 PK, Bilbao 48080, Basque Country, Spain }}

\maketitle
\begin{abstract}
We derive the linearly perturbed matching conditions
between a Schwarzschild spacetime region with stationary and axially symmetric
perturbations and a FLRW spacetime with arbitrary perturbations. The
matching hypersurface is also perturbed arbitrarily and, in all cases,
the perturbations are decomposed into scalars using the Hodge operator
on the sphere.  This allows us to write down the matching conditions
in a compact way.
In particular, we find that the existence
of a perturbed (rotating, stationary and vacuum)
Schwarzschild cavity in a perturbed FLRW universe forces the
cosmological perturbations to satisfy constraints that link
rotational and gravitational wave perturbations. 
We also prove that if the perturbation on the
FLRW side vanishes identically, then the vacuole must be
perturbatively static and hence Schwarzschild. 
By the dual nature of the problem, the first result translates into
links between 
rotational and gravitational wave perturbations on a perturbed
Oppenheimer-Snyder model, where the perturbed FLRW dust collapses
in a perturbed Schwarzschild environment which rotates in equilibrium.
The second result
implies in particular that no region described by FLRW can be a source of the
Kerr metric.

\end{abstract}

\section{Introduction}
A long standing question in cosmology concerns the way large scale dynamics 
influences the behaviour on smaller scales. 
The most common view regarding this question is that the influence of
the cosmic expansion on local physics is zero or negligible. The main
argument supporting this conclusion is based on the Einstein-Straus
model \cite{Einstein-Straus} which consists of a vacuum
spherical cavity (described by the Schwarzschild metric, hence static)
embedded in an expanding dust Friedmann-Lema\^itre-Robertson-Walker
(FLRW) model. The matching between both spacetimes is performed across
a timelike hypersurface using the standard matching theory in general
relativity, usually known as the Darmois matching conditions (we refer
to \cite{Mars-Seno} for a full account on matching general
hypersurfaces). In this model the local
physics occurs inside the Schwarzschild vacuole which, being static,
perceives no effect of the cosmological expansion.

Despite its clear physical interpretation, this model presents serious
problems and involves a number of idealisations which include a
spatially homogeneous and isotropic cosmological model, the assumption
of spherical symmetry both for the metric inside the vacuole and for
its boundary, and the assumption of an exact  static vacuum in the
interior. Indeed, more sophisticated models have been constructed
by using Lema\^itre-Tolman-Bondi regions for the cosmological part
(see \cite{KR,Bonnor00}), and other type of cavities \cite{FJLS,SenoComp},
but all these models are spherically symmetric.   
An important question is whether the
Einstein-Straus' conclusion is robust with respect to various
plausible (non-spherically symmetric) generalisations.

A general property of the matching theory
is that solving a matching problem always gives, as
an immediate consequence, that the ``complementary''  matching 
also holds \cite{SenoComp}.
Consequently, the `a priori' interior and exterior roles
assigned to the Schwarzschild and FLRW regions in the Einstein-Straus
model can be interchanged,
and all the previous results also apply to the
Oppenheimer-Snyder model of collapse \cite{OPSN},
in which a spherically symmetric
FLRW region of dust collapses in a Schwarzschild exterior geometry. 

The first attempts to provide non-spherically symmetric
generalisations can be found in the works by Cocke in \cite{Cocke1}
and Shaver \& Lake in \cite{ShLa}.
The first conclusive discussion of different metrics and shapes of the static
region appeared
in a paper by Senovilla \& Vera \cite{Seno-Vera}, where it was shown
that a \emph{locally}
cylindrically symmetric static region cannot be matched to an
expanding FLRW model across a non-spacelike
hypersurface preserving the cylindrical
symmetry, irrespective of the matter content in the
cylindrically symmetric region.  Therefore, a
detailed analysis became necessary in order to decide whether the
assumption of spherical symmetry of the static region was a fundamental
ingredient for the models.
This analysis was performed by Mars, who showed in two steps, 
\cite{Mars98} and \cite{Mars01}, that
a static region matched to a FRLW cosmological model is 
forced to be spherically symmetric (both in shape and 
spacetime geometry) under very weak conditions
on the matter content, which include vacuum as
particular case. Therefore, \textit{the only static vacuum region that
can be matched to a non-static FLRW is an (either interior or
exterior) spherically shaped region of Schwarzschild}, which thus
leads to the Einstein-Straus or the Openheimer-Snyder models.

To summarise, the Einstein-Straus model, consisting of a static vacuole
embedded in an exact FLRW geometry (necessarily of dust),
does not allow any non-spherical generalisation and, in this sense,
it is unstable. The same applies to the Oppenheimer-Snyder model.

Regarding the assumption of staticity, it has recently
been shown by Nolan \& Vera \cite{Nolan-Vera} that if a stationary
and axially symmetric region is to be matched to a non-static FLRW region
across a hypersurface preserving the axial symmetry, then the stationary
region must be static. Hence, the results in \cite{Mars98,Mars01} can
be applied to any stationary and axisymmetric region.
In particular, \textit{the only stationary and axisymmetric vacuum
region that can be matched to FLRW preserving the axial symmetry is an
(either interior or exterior) spherically shaped region of
Schwarzschild.}  This implies that the Einstein-Straus and
Oppenheimer-Snyder models cannot be generalized by including
stationary rotation in the non-FLRW region.  Note that this shows in
particular that no FLRW axially symmetric region can be a source of
Kerr.

Now, two possibilities in trying to generalise such models can be
considered: the first is to take non-spherical exact solutions
which generalise the FLRW region (such as Bianchi models) and the
second to consider non-spherical perturbations of FLRW. The first
possibility was considered by Mena, Tavakol \& Vera in \cite{MTV}, 
where they studied a matching preserving the symmetry of a
cylindrically symmetric interior spacetime with locally rotationally
symmetric spatially homogeneous (but anisotropic) exteriors. This
matching resulted in restrictive generalisations of the
Einstein-Straus model \cite{MTV}, none of them physically
admissible.  Therefore it is interesting
to consider the second possibility, i.e. to construct a perturbed
model.

Perturbed matching conditions have been applied many times in the
past: Hartle \cite{Hartle} studied first and second order stationary
and axisymmetric rigidly rotating perturbations of static
perfect-fluid balls in vacuum. Chamorro performed the first order
matching of a Kerr cavity
in an expanding perturbed FLRW model \cite{Chamorro}.
For spherical symmetry,
the linearised matching conditions in an arbitrary gauge were studied
by Gerlach \& Sengupta \cite {gersen_oddmatch,gersen_evenmatch}
and by Mart\'in-Garc\'ia \& Gundlach \cite{josemcarsten}.  There are also
studies by Cunningham, Price \& Moncrief where axial \cite{CPM78} and
polar \cite{CPM79} perturbations of the Oppenheimer-Snyder model of
collapse were derived.
However, a matching perturbation theory in general relativity has only
recently been developed in its full generality for first order
\cite{Battye-Carter, Mukohyama} and second order perturbations
\cite{Mars05}.
A critical review about the study of linear perturbations
of matched spacetimes including gauge problems has been recently presented
in \cite{MMV1}.

Another interesting approach has been followed by
Dole\v{z}el, Bi\v{c}\'ak and Deruelle \cite{voidrotshell},
who have studied slowly rotating
voids in cosmology with a model consisting on
an interior Minkowskian void, a matter shell between
the void and the cosmological model and a FLRW universe
with a particular type of perturbation describing rotation.
We emphasise that, in this paper, we focus on
generalisations of the Einstein-Straus and the Openheimer-Snyder
models without surface layers of matter (i.e. such that
the Darmois matching conditions are satisfied),
and we do not restrict the FLRW perturbations in any way.

In this paper we consider first order (linear) perturbations of the
Einstein-Straus model, as well as perturbations of the
Oppenheimer-Snyder model (as the perturbative matching satisfies the
same dual property as the full matching).
Our background model consists then of a
Schwarzschild region matched to a FLRW dust cosmological model across
a spherically symmetric (timelike) hypersurface.
We perturb the Schwarzschild
part with vacuum stationary axially symmetric perturbations and we perturb
the matching surface as well as the exterior FLRW region with
arbitrary perturbations, not restrained to any material content.

Our approach to this problem consists in exploiting the underlying spherical
symmetry of the background and of the matching hypersurface as much as
possible. This is normally done in the literature by resorting to
decompositions of all objects in terms of scalar, vector and
tensor harmonics on the sphere. Our aim is to use an alternative
method based on the Hodge decomposition of all tensor objects on the sphere
in terms of scalars. The two approaches are obviously related to each 
other. However, by working with Hodge scalars we avoid the need to 
deal  with
infinite series of objects (one for each $l$ and $m$ in the spherical
harmonic decomposition). In particular, our set of matching conditions
has a finite number of equations (involving scalars that depend on the
three coordinates in the matching hypersurface) instead of an infinite
collection of matching conditions for functions of only one variable (the
time coordinate on the matching hypersurface). The equations are therefore 
much more compact. 

In fact, even when working with $S^2$ scalars the 
length of the equations can grow substantially depending on how they
are combined and written down. This is partly due to the
level of generality we leave for the matching hypersurface and the FLRW
perturbations. We have taken the effort
to combine the equations and to group several terms in each equation
in such a way that the set of equations becomes reasonably short and 
manageable. We emphasize that the gauge in the FLRW part will be left
completely free, so that the readers may choose their favourite one.
As an example, we rewrite the equations in the
Poisson gauge in Appendix \ref{sec:poisson_gauge}.

Deriving and writing down the linearised matching conditions for
our perturbation of the Einstein-Straus (and Oppenheimer-Snyder)
model is the main result of this paper.
A more detailed analysis of the resulting set is postponed to a later paper.
However, in order to show the usefulness and power of the equations, we present
two applications. The first one is based on the observation that 
the linearised matching conditions can be combined in such a way that
two equations {\it involving only terms on the FLRW side} hold.
These equations are therefore constraints on the
perturbations in the FLRW part that must be necessarily satisfied
if an interior stationary and axisymmetric perturbed vacuole is 
present (and hence the local physics can remain unaffected by the cosmic
expansion, as generally believed). 
These two constraints link the vector and tensor
(linear) FLRW perturbations on the boundary of the region,
and they imply, basically, that \textit{if the perturbed FLRW
region contains vector modes (with $l\geq 2$ harmonics)
on the boundary, then it must also contain tensor modes.}
In other words, if a (perturbed) FLRW region contains rotational perturbations
that reach the boundary of a perturbed stationary and axially symmetric
vacuum region, then the cosmological model there must also
carry gravitational waves. Given the estimates of cosmological rotational
perturbations through observations
(see e.g. \cite{cosm_rot} and references therein), the constraints due to the
possible existence of stationary and axisymmetric vacuoles would
provide estimates of cosmological gravitational waves.

As a second application, we consider the case when the FLRW part of the
spacetime remains exact (i.e. all perturbations vanish there). 
As discussed above, the 
results in \cite{Nolan-Vera} combined with \cite{Mars01}
imply that the interior has to be exactly Schwarzschild
{\it provided the matching hypersurface is assumed to be axially
symmetric}. Since we allow for non-axially symmetric perturbations
of the matching hypersurface we can address the question of whether this
result generalizes to arbitrary hypersurfaces (at the linear level,
of course). Our conclusion is that indeed this is the case. This
result points, once again, into the fragility of the Einstein-Straus
model against any reasonable generalisation. From the
Oppenheimer-Snyder model point of view, this also means that a body
modeled by a dynamical FLRW model, irrespective of its shape and its
relative rotation with the exterior, cannot be the source of any
stationary (non-static), axially symmetric vacuum exterior, in
particular of the Kerr metric.

The paper is organised as follows. In Section 
\ref{Matching_in_brief} 
we give a very brief
summary of the linearised matching theory, where we fix our
notation. In Section \ref{spherical-match}
we summarize the
standard theory of the Hodge
decomposition of vectors and symmetric tensors on the sphere
and apply the theory to first order perturbations of a spherical
background in general. The use of Hodge decompositions introduces
a so-called {\it kernel freedom}  which is discussed and analysed.
Section \ref{interior} is devoted to obtaining the explicit form
of vacuum, stationary and axially symmetric perturbations of the
Schwarzschild region. After perturbing the background matching 
hypersurface arbitrarily, we obtain, in subsection \ref{intscalars},
the explicit form of the
Hodge scalars of the perturbed
first and second fundamental forms of the matching hypersurface.
The same procedure is followed in Section \ref{exterior} for the FLRW side.
Section \ref{MainSection} is devoted to obtaining the linearised
matching conditions for our problem in terms of $S^2$ scalars. As in 
any Hodge decomposition, the set of equations decompose into 
odd and even equations, and both sets are carefully combined and
rewritten to make them as compact as possible. 
The equations given in this section constitute the main
result of this paper, so we summarize the hypotheses and
conclusion in Theorem \ref{MainTheorem}. Sections \ref{sec:constraints} 
and \ref{ExactFLRWPertSch}
contain the two applications we present in this paper. Section
\ref{sec:constraints}
is devoted to deriving the constraints in the FLRW side 
and Section \ref{ExactFLRWPertSch}
 to the matching with an exact FLRW. 
The paper contains four Appendices. Appendices \ref{appendixA}
 and \ref{appendixB}
contain, respectively for the Schwarzschild and
the FLRW parts, the full expressions for the linear perturbations
of the first and second fundamental forms prior to their Hodge
decomposition. Appendix \ref{sec:poisson_gauge} specifies
our general matching conditions
in the Poisson gauge in the FLRW part. Finally, 
Appendix \ref{variable-identif} is devoted to the comparison
between the expressions used in our formalism and
the doubly gauge invariant quantities in \cite{Mukohyama} and \cite{MMV1}.

Lower case Latin indices at the beginning of the alphabet
$a,b,...= 1,2,3$ refer to tensors on the
constant cosmic time hypersurface in FLRW, at the middle
of the alphabet, $i,j,...= 1,2,3$ are used for tensors on the
matching hypersurface.  
The first upper case Latin indices 
$A,B,...= 2,3$ denote tensors on the sphere while
middle indices $I,J,...=0,1$ are used for tensors in the 
surfaces orthogonal to the spherical orbits.
Finally, Greek indices
$\alpha,\beta,...=0,1,2,3$ refer to general spacetime tensors.

\section{Linearised perturbed matching theory in brief}
\label{Matching_in_brief}
The linearised matching involves perturbing a background which
is already constructed from the matching of two regions 
$(M^{+}_0,\gback{}^{+})$
and $(M^{-}_0,\gback{}^{-})$,
with corresponding boundaries $\supo^{\pm}$
which are diffeomorphic to each other (thence identified
as the matching hypersurface).
Taking local coordinates on the matching hypersurface amounts to write
down two embeddings
\begin{eqnarray}
\label{embed}
\Phi_\pm:\; \supo &\longrightarrow& M^\pm_0\\
\label{eq:embeddings}
             \xi^i &\mapsto& x_\pm^\alpha=\Phi_\pm^\alpha(\xi^i),
\nonumber
\end{eqnarray}
such that $\Phi_\pm(\supo) = \supo^{\pm}$, where $x^{\alpha}_\pm$ are 
arbitrary coordinates on each of the background regions. 
The coordinate vectors $\partial_{\xi^i}$ intrinsic to $\supo$
define tangent vectors to the boundaries $
e^{\pm \alpha}_i = \frac{\partial \Phi_\pm^\alpha}{\partial \xi^i}$.
Assuming $\Sigma_0^{\pm}$ not to be null at any point, there is
a unique up to orientation unit normal vector $\nback_{\pm}{}^{\alpha}$. 
The orientation of one of the normals can be chosen arbitrarily but
the other must be chosen accordingly so that both point to the same
region after the matching.
The first and second fundamental forms are, respectively, 
$\fffback{}^{\pm}_{ij}\equiv e^{\pm \alpha}_i e^{\pm \beta}_j
\gback{}^{\pm}_{\alpha\beta}|_{\supo^\pm}, \quad
\sffback{}^{\pm}_{ij}=-\nback_{\pm\alpha}
e^{\pm \beta}_i\nabla^\pm_\beta e^{\pm \alpha}_j|_{\supo^\pm}$
and the background matching conditions require\footnote{
As mentioned earlier, we are not interested in a resulting spacetime with a
non-vanishing energy-momentum tensor with support on the matching
hypersurface (a ``shell''), and therefore
we do not admit jumps in the second fundamental form.
}
\begin{equation}
  \fffback_{ij}^{+}=\fffback_{ij}^{-},~~~
  \sffback_{ij}^{+}=\sffback_{ij}^{-}.
\label{eq:backmc}
\end{equation}

Consider now a perturbation of the background metric
$
g^\pm_{pert}={\gback}{}^\pm+{\gfpert}{}^\pm
$ 
and of the boundaries
$\supo^\pm$ via the vector fields $\vec Z^\pm=Q^\pm \vnback_\pm+
\vec T^\pm|_{\supo^\pm}$, where $\vec T^\pm$ are tangent to $\supo^\pm$.
The linearised matching conditions are derived in
\cite{Battye-Carter} and \cite{Mukohyama} 
 (see also \cite{Mars05} for second order matching),
and read
\begin{equation}
\ffffpert{}^+_{ij} = \ffffpert{}^-_{ij} ,~~~ 
\sfffpert{}^+_{ij} = \sfffpert{}^-_{ij} ,
\label{eq:fpertmc}
\end{equation}
with (for a timelike matching hypersurface)
\begin{eqnarray}
\label{pertfirst}
\ffffpert{}^\pm_{ij}&=&
{\cal L}_{\vec T^\pm} \fffback{}^{\pm}_{ij}+
2 Q^\pm \sffback{}^{\pm}_{ij}+
e^{\pm\alpha}_i e^{\pm\beta}_j \gfpert{}^\pm_{\alpha\beta}|_{\supo^\pm},\\
\label{pertsecond}
\sfffpert{}^\pm_{ij}&=&
{\cal L}_{\vec T^\pm} \sffback{}^{\pm}_{ij}-D_iD_j Q^\pm+
Q^\pm(\nback_\pm{}^\mu \nback_\pm{}^\nu
R^{(0)\pm}_{\alpha\mu\beta\nu}e^{\pm\alpha}_i e^{\pm\beta}_j+
\sffback{}^{\pm}_{il}\sffback^{l}{}^{\pm}_j)
\nonumber\\
&+&\frac{1}{2}\gfpert{}^\pm_{\alpha\beta}
\nback_\pm{}^\alpha \nback_\pm{}^\beta \sffback{}^{\pm}_{ij}-
\nback_\pm{}_\mu S^{(1)\pm\mu}_{\alpha\beta} e^{\pm\alpha}_i e^{\pm\beta}_j|_{\supo^\pm},
\end{eqnarray}
where 
$D_i$ is the three dimensional covariant derivative of
$(\supo,\fffback{}^{\pm}_{ij})$
and
\[
S^{(1)\pm\alpha}_{\beta\gamma}\equiv
\frac{1}{2}\left(
\nabla^{\pm}_\beta\, \gfpert{}^{\pm\alpha}_\gamma+
\nabla^{\pm}_\gamma\, \gfpert{}^{\pm\alpha}_\beta-
\nabla^{\pm \, \alpha}\, \gfpert{}^\pm_{\beta\gamma}
\right).
\]
The tensors $\ffffpert{}^{\pm}$ and $\sfffpert{}^{\pm}$ are
spacetime gauge 
invariant by construction, since they are objects
intrinsically defined on $\supo^\pm$,
and therefore conditions (\ref{eq:fpertmc}) are spacetime gauge 
invariant.
Moreover, it turns out that the equations (\ref{eq:fpertmc})
are also hypersurface gauge invariant
provided the background is properly matched
(i.e. once (\ref{eq:backmc}) hold).

The quantities $Q^\pm$ and $\vec T^\pm$ are unknown \textit{a priori},
and fulfilling the matching conditions requires \textit{showing}
that two vectors $\vec Z^\pm$ exist such that (\ref{eq:fpertmc})
are satisfied.
The spacetime gauge freedom can be used to fix either or both
vectors $\vec Z$, but these choices have to be avoided, or carefully analysed,
if additional spacetime gauge choices are made, as otherwise
the possible matchings might be restricted artificially.
On the other hand, the hypersurface gauge can be used
in order to fix one of the vectors $\vec T^+$ or $\vec T^-$, but not both
(see  \cite{MMV1} for a full discussion). 
\section{Hodge decomposition of the linearised perturbed matching in
spherical symmetry}
\label{spherical-match}
From now onwards we shall concentrate on spherically symmetric
background configurations.  In order to write the matching conditions
in a way which exploits the symmetries we shall
use the Hodge decomposition on the sphere. There are some good reasons
to do that. As outlined in the Introduction, the equations naturally
inherit the spherical symmetry of the background configuration,
and thus one expects that any approach based on spherical decompositions will
render the equations in a simpler form.
Previously in the literature, this has been implemented by decomposing
the relevant quantities
into scalar, vector and tensor harmonics (see e.g. \cite{Mukohyama}). 
From a formal point of view,
those decompositions are very useful since
they provide independent sets of equations for the different
spectral values, say $l$ and $m$. However, in practice,
the harmonic decomposition may become a problem on its own
when studying explicit models.
Moreover, even if the problem can be formally solved for the infinite
spectrum, the sum convergence of the resulting decomposition
should be eventually ensured.

Using the Hodge decomposition has the advantage that one works with all
the different $(l,m)$ harmonics of a given quantity at once. In fact, the Hodge
scalars correspond in a suitable sense, to the resummation of the 
previous spectral decomposition. For instance instead
of using all the (infinite number of) equations that correspond to
each value of $l$ and $m$ in the matching equations involving
$(\sigma_{(L)0})_{lm}$ (see \cite{Mukohyama}),
only one equation for the whole sum
$\sum_{lm} (\sigma_{(L)0})_{lm} Y_l^m$ ($\equiv F$ here) is needed.

It is clear that one can always 
go from the Hodge scalars to the spherical harmonics decomposition in a
straightforward way. However, it is not always easy to rewrite the
infinite number of expressions appearing in a spectral decomposition in 
terms of Hodge scalars. Working with Hodge scalars involves  a finite number
of equations and obviously there arise no convergence problems 
(although then one often has to deal with 
PDE in 3+1 dimensions instead of 1+1 equations, which are simpler).
Furthermore, their calculation entails a quite straightforward procedure,
more easily implemented in algebraic computing.
We devote the Appendix \ref{variable-identif} at the end of the
paper to relate the functions of the Hodge decomposition used
in the present paper and the coefficients used in \cite{Mukohyama,MMV1}
for the scalar, vector and tensor harmonic decomposition.

\subsection{The Hodge decomposition on the sphere}
\label{kernels}
We recall that the Hodge decomposition on $S^2$ tells
us that any one-form $\bm{V}$ on $S^2$ can be canonically decomposed as
\begin{equation}
\bm{V} = d F + \star dG, \label{Hodge}
\end{equation}
where and $F$ and $G$ are functions on the sphere and $(\star
dG)_A= \eta^{C}{}_{A}D_C G $
is the Hodge dual with
respect to the round unit metric $h_{AB}dx^A dx^B=d \vartheta^2 + \sin^2
\vartheta d \varphi^2$ on $S^2$. The corresponding volume form
and covariant derivative are denoted respectively by
$\eta_{AB}$ and $D_A$. Latin indices
$A,B, ...$ are raised and lowered with $h_{AB}$ and the orientation is chosen so that 
$\eta_{\vartheta \varphi} >0$.
Furthermore, any symmetric tensor $T_{AB}$ on the sphere can be canonically
decomposed as 
\[
T_{AB} = D_A U_B + D_B U_A + H h_{AB},
\]
for some 1-form $U_A$ on $S^2$, which, in turn,
can be decomposed as
$
U_A = D_A P +  (\star d R )_A.
$

The Hodge decomposition on the sphere has a non trivial kernel, i.e. the
zero vector and the zero tensor on $S^2$ can be decomposed in
terms non-vanishing scalars, albeit of a very special form. 
First, we consider the kernel corresponding to the
vanishing vector on the sphere:
\[
D_A F + \eta^B{}_{A} D_B G = 0.
\]
Since the only harmonic functions on the sphere are the constant functions,
it follows that $F$ and $G$ must be independent of the angular coordinates
$\{x^A\}$($=\{\vartheta,\varphi\}$).
Regarding the zero symmetric tensor $T_{AB}$, we must solve
\begin{equation}
D_A U_B + D_B U_A + H h_{AB} = 0,
\label{eqAB}
\end{equation}
which states that $U^A$ is a conformal
Killing vector on the sphere.
There are six conformal Killing vectors on the sphere: three
proper ones and three Killing vectors.
They correspond to the usual
longitudinal and transverse
$l=1$ vector  harmonics  respectively
(denoted as $V_{(L)}{}^A$ and $V_{(T)}{}^A$
with $l=1$ in \cite{Mukohyama,MMV1}).
Explicit expressions for the conformal Killing vectors
are obtained using the $l=1$ spherical harmonics $Y^m_1$ ($m=1,2,3$)
\begin{eqnarray*}
Y_1^1 = \cos \vartheta, \hspace{1cm}
Y_1^2 = \sin \vartheta \cos \varphi, \hspace{1cm}
Y_1^3 = \sin \vartheta \sin \varphi.
\end{eqnarray*}
The gradients $D_A Y^m_1$ correspond to
three linearly independent proper conformal Killing vectors on $S^2$,
and their Hodge duals
$\eta^B{}_A D_B Y^m_1$ correspond to three linearly
independent Killing vectors on $S^2$.
Therefore, decomposing further $U_A$
as $U_A = D_A P + \eta^B{}_A D_B R$ 
we obtain
\begin{eqnarray*}
P &=& P_0 + \sum_{m} P_m  Y^m_1\\
R &=& R_0 + \sum_{m} R_m  Y^m_1
\end{eqnarray*}
for some eight free coefficients $P_0$, $P_m$, $R_0$, $R_m$
independent of $\{\vartheta,\varphi\}$.
Substitution into (\ref{eqAB}) leads to 
\[
H = 2  \sum_{m} P_m Y^m_1.
\]

\subsection{Perturbed matching conditions in terms of scalars}
\label{sec:scalar-match}

Our background configuration is spherically symmetric and
composed of two 
spherically symmetric spacetimes $(M_0^\pm,{\gback}{}^{\pm})$
matched across spherically symmetric timelike boundaries $\supo^\pm$
diffeomorphic to each other.

Let us for definiteness concentrate on the $(M_0^+,\gback{}^{+})$ spacetime
and drop the $+$ subindex
(analogous expressions obviously hold for the $(M_0^-,{\gback}{}^{-})$
spacetime region). We choose coordinates adapted to the spherical symmetry,
so that
\[
\gback_{\alpha\beta}dx^\alpha dx^\beta=\omega_{IJ}dx^I dx^J+r^2(x^I)
(d\theta^2+\sin^2\theta d\phi^2),
\]
where $\omega_ {IJ}$ is a Lorentzian two-dimensional metric
and $r(x^I)  \geq 0 $. A general spherically symmetric boundary 
can be described by the embedding (or parametric form)
\begin{equation}
\supo:=\{x^0=\Phi^0_{(0)}(\lambda),x^1=\Phi^1_{(0)}(\lambda),\theta=\vartheta,
\phi=\varphi\},
\label{eq:gen_sph_supo}
\end{equation}
where $\{\xi^i\}=\{\lambda,\vartheta,\varphi\}$
is a coordinate system in $\supo$ adapted to the spherical symmetry.

The coordinate tangent vectors to $\supo$ read
\begin{equation}
\vec e_\lambda=\dot\Phi^0_{(0)}\partial_{x^0}
+\dot\Phi^1_{(0)}\partial_{x^1}|_{\supo},~~
\vec e_{\vartheta}=\partial_\theta|_{\supo},~~
\vec e_{\varphi}=\partial_\phi|_{\supo},
\label{eq:gen_es}
\end{equation}
where the dot denotes a derivative w.r.t. $\lambda$.
Defining $N^2\equiv -e_\lambda{}^I e_\lambda{}^J\omega_{IJ}|_{\supo}$
the unit normal to the boundary reads
\begin{equation}
\fnback=\frac{\sqrt{-\det \omega}}{N}(-\dot\Phi^1_{(0)} dx^0
+ \dot\Phi^0_{(0)} dx^1)|_{\supo}.
\label{eq:gen_n}
\end{equation}
The sign of $N$ corresponds to the choice of orientation of the normal.
The first and second fundamental forms on $\supo$ read
\begin{eqnarray}
  \label{eq:fffback}
  \fffback_{ij}d\xi^i d\xi^j=-N^2 d\lambda^2+r^2|_{\supo} \domegas, \\
  \label{eq:sffback}
  \sffback_{ij}d\xi^i d\xi^j=N^2\K d\lambda^2+
  r^2|_{\supo}\bar\K\domegas,
\end{eqnarray}
where
$
\K\equiv N^{-2}e_\lambda{}^I e_\lambda{}^J\nabla_I \nback_J|_{\supo},
\bar\K=\nback{}^I\partial_{x^I}\ln r|_{\supo}.
$

Applying these expressions to each spacetime region 
$(M^{\pm}_0,\gback{}^{\pm})$,
the background matching conditions (\ref{eq:backmc}) become
\begin{equation}
  \label{eq:backmc2}
  N^2_+=N^2_-,~~~ r_+|_{\supo}=r_-|_{\supo},~~~
  \K_+=\K_-,~~~ \bar\K_+=\bar\K_-.
\end{equation}

These equations involve scalars on the sphere and therefore
do not require any further Hodge decomposition.
Notice that the embeddings have been implicitly chosen so that
$\theta_+=\theta_-(=\vartheta)$ and
$\phi_+=\phi_-(=\varphi)$ on the matching hypersurface.
This can in principle be modified 
by an arbitrary rigid rotation, which is an intrinsic freedom
of any matching preserving
the symmetry \cite{mps}. At the background level, this rigid rotation
is irrelevant and can be re-absorbed by a coordinate change. However,
its effect is not so trivial at the perturbed level (see 
\cite{MMV1} for a discussion on its consequences).

Consider now an arbitrary linear perturbation and use the
Hodge decomposition applied to the
perturbed first and second fundamental forms. More specifically, we write
\begin{eqnarray}
  \ffffpert{}^\pm_{\lambda A} &=&
  D_A F^q_\pm + ( \star d G^q_\pm)_{A}\nonumber\\
  \sfffpert{}^\pm_{\lambda A} &=&
  D_A F^k_\pm + (\star d G^k_\pm )_A\nonumber\\
  \ffffpert{}^\pm_{AB} &=&
  D_A \left ( D_B P^q_{\pm} + (\star d R^q_{\pm})_B \right )
+  D_B \left ( D_A P^q_{\pm} + (\star d R^q_{\pm})_A \right )
+ H^q_\pm h_{AB}\label{DefsS2Scalars}\\
  \sfffpert{}^\pm_{AB} &=&
  D_A \left ( D_B P^k_{\pm} + (\star d R^k_{\pm})_B \right )
+  D_B \left ( D_A P^k_{\pm} + (\star d R^k_{\pm})_A \right )
+ H^k_\pm h_{AB}\nonumber
\end{eqnarray}
where 
$F^q_\pm$, $G^q_\pm$, $P^q_\pm$, $R^q_\pm$, $H^q_\pm$,
$F^k_\pm$, $G^k_\pm$, $P^k_\pm$, $R^k_\pm$, $H^k_\pm$,
are scalar functions on $S^2$ that depend on the parameter $\lambda$.
The linearised matching conditions (\ref{eq:fpertmc})
can  be rewritten as conditions involving 
$\ffffpert{}_{\lambda\lambda}^\pm$ and $\sfffpert{}_{\lambda\lambda}^\pm$
together with the functions above. Recalling the existence
of a non-trivial kernel for the Hodge decomposition, the 
equalities in (\ref{eq:fpertmc}) turn out to be equivalent to 
\begin{eqnarray}
\label{eq:match-scalars}
\ffffpert{}^-_{\lambda \lambda}&=&
\ffffpert{}^+_{\lambda \lambda} \nonumber\\
F^q_- &=& F^q_+ - N^2 {\markkern \Fqo (\lambda)} , \nonumber\\
G^q_- &=& G^q_+ - N^2 {\markkern\Gqo  (\lambda)},  \nonumber\\
P^q_- &=& P^q_+ - N^2 {\markkern  
\left ( \Pqo  (\lambda)+ \Pqm  (\lambda) Y^m_1 \right )},  \nonumber\\
R^q_- &=& R^q_+ -  N^2 {\markkern  
\left ( \Rqo  (\lambda) + \Rqm  (\lambda) Y^m_1 \right )},  \nonumber\\
H^q_- &=& H^q_+ - 2  N^2 {\markkern\Pqm  (\lambda)Y^m_1}, \\
\sfffpert{}^-_{\lambda \lambda}&=&
\sfffpert{}^+_{\lambda \lambda} \nonumber\\
F^k_- &=& F^k_+ -  N^2 {\markkern\Fexto  (\lambda)},  \nonumber\\
G^k_- &=& G^k_+ -  N^2 {\markkern\Gexto  (\lambda)},  \nonumber\\
P^k_- &=& P^k_+ -  N^2 {\markkern  
\left ( \Pexto  (\lambda) + \Pextm  (\lambda) Y^m_1 \right )},  \nonumber\\ 
R^k_- &=& R^k_+ -  N^2 {\markkern  
\left ( \Rexto  (\lambda) + \Rextm  (\lambda) Y^m_1 \right ) },  \nonumber\\
H^k_- &=& H^k_+ - 2  N^2 \markkern\Pextm{  (\lambda) Y^m_1}, \nonumber
\end{eqnarray}
where all the functions  with a $0$ or $m$ subindex (and a different
color and size) depend only on $\lambda$
and correspond to the kernel freedom discussed above.
They will collectively be named
as {\it kernel functions} in what follows. The explicit factor
$N^2(\lambda)(\equiv N_+^2=N_-^2)$ in front of these functions has
been added for convenience, as it simplifies some of the expressions
below.

It may seem that adding these kernel functions is redundant,
as they do not affect the
tensors $\ffffpert{}^{\pm}_{ij}$ and $\sfffpert{}^{\pm}_{ij}$.
However, it is precisely the fact
that we want to impose the matching conditions at the level
of $S^2$ scalars that forces us to include them. 
From a practical point of view, the explicit inclusion
of the kernel functions in equations (\ref{eq:match-scalars})
allows one to choose arbitrarily
any particular decomposition at either ($\pm$) side.
In particular, when studying existence problems for the matching of two 
given configurations (decomposed in terms of $S^2$ scalars in an explicit manner) it is important to keep 
the kernel functions free, as they may serve to fulfill conditions which 
might otherwise seem to be incompatible.
For an explicit case where the kernel functions
turn out to be relevant, see Section \ref{ExactFLRWPertSch} below.

Summarizing, equations (\ref{eq:match-scalars}) are the formal linearly
perturbed matching conditions written in terms of $S^2$ scalars.
The next task is to evaluate explicitly all the scalars
involved in the Hodge decomposition of
(\ref{pertfirst}) and (\ref{pertsecond}) in the cases
we will be considering, namely
the matching of a Schwarzschild spacetime with a stationary and axially symmetric vacuum 
linear perturbation and a FLRW spacetime with a general linear perturbation. 
\section{Perturbed Schwarzschild region $(-)$}
\label{interior}
First, we describe the perturbations of the Schwarzschild
region (denoted by the ($-$) sign)
and derive the perturbed first and second fundamental forms on
$\supo^-$.
\subsection{Stationary and axially symmetric perturbations of Schwarzschild}
\label{intpert}
We start by taking a stationary and axially symmetric vacuum metric which in 
Weyl-Papapetrou coordinates can be written
as \cite{sol}
\begin{equation}
\label{staticWeyl}
\gback{}^{-}  = -e^{2 U} (dt+A d\phi)^2 + e^{-2 U} \left [ e^{2 k} \left ( d\rho^2
+ d z^2 \right ) + \rho^2 d\phi^2 \right ],
\end{equation}
where $U, A$ and $k$ are functions of $\rho$ and $z$. 
Vacuum linear perturbations can be obtained by taking derivatives of (\ref{staticWeyl})
with respect to a perturbation parameter. The result is
\begin{eqnarray}
\label{Weylform}
\gfpert{}^{-}&=& -2 e^{2 U} U^{(1)} dt^2 -4 U^{(1)} A e^{2U} d\phi dt
-2U^{(1)}e^{2U}A^2d\phi^2- 2 e^{2 U} A^{(1)} dt d\phi  \nonumber\\
&-&2A A^{(1)}e^{2U}d\phi^2 + 2 e^{- 2 U}
e^{2k} \left ( - U^{(1)} + k^{(1)} \right ) \left (d\rho^2 + d z^2 \right )  
- 2 e^{-2 U} U^{(1)} \rho^2 d\phi^2,
\end{eqnarray}
where the perturbation is obviously written in a specific gauge, which 
we shall denote by {\it Weyl gauge}. The functions
$U^{(1)}$, $A^{(1)}$ in (\ref{Weylform}) 
depend  on $\rho$ and $z$ and satisfy the perturbed vacuum equations, written
explicitly below.

The Schwarzschild background is obtained from (\ref{staticWeyl}) by setting
\[
U=\frac{1}{2}\log \left(1-\frac{2 m}{r}\right), \quad e^{2 k} =
\frac{r \left (r - 2m \right )}{(r-m)^2 - m^2 \cos^2 \theta}, \quad A=0,
\]
where
\[
\rho = r \sin \theta \sqrt{ 1 - \frac{2m}{r} }, \quad z = \left (r - m
\right ) \cos \theta,
\]
so that the background metric reads
\[
\gback{}^{-}=- \left (1- \frac{2 m}{r} \right ) dt^2 + \frac{dr^2}{1 - \frac{2 m}{r}}
+ r^2 \left ( d\theta^2 + \sin^2 \theta d \phi^2 \right ).
\]
In these coordinates, the first order perturbation metric (\ref{Weylform})
becomes
\begin{eqnarray*}
\gfpert{}^{-}&=& 
- 2 \left (1- \frac{2 m}{r} \right ) 
\left ( U^{(1)} dt^2+   {A^{(1)}} dt d \phi \right )\\
&&- 2 r^2 \sin^2 \theta  {U^{(1)}} d\phi^2 
+ 2 \left (k^{(1)} - U^{(1)} \right ) \left ( \frac{dr^2}{1 - \frac{2 m}{r}}
+ r^2 d \theta^2 \right ).
\end{eqnarray*}
The perturbed vacuum equations decompose 
into a pair of decoupled second order PDE
for $U^{(1)}$ and $A^{(1)}$ 
\begin{eqnarray}
r \left ( r - 2 m \right ) \frac{\partial^2 {U^{(1)}}}{\partial r^2} 
+ \frac{\cos \theta}{\sin \theta}  \frac{\partial
{U^{(1)}}}{\partial \theta} 
+ \frac{\partial^2 {U^{(1)}}}{\partial \theta^2}
+ 2 \left ( r - m \right ) \frac{\partial
{U^{(1)}}}{\partial r} &=& 0, \label{equationUprime} \\ 
r \left ( r - 2 m \right ) \frac{\partial^2 {A^{(1)}}}{\partial r^2}
- \cs \frac{\partial {A^{(1)}}}{\partial \theta}
+ \frac{\partial^2 {A^{(1)}}}{\partial \theta^2}
- 4 m \frac{\partial {A^{(1)}}}{\partial r} &=& 0, \label{equationAprime}
\end{eqnarray}
together with a first order system for $k^{(1)}$ (which is
compatible provided (\ref{equationUprime}) holds)
\begin{eqnarray*}
\frac{\partial {k^{(1)}}}{\partial r} &=&
\frac{2 m \sin \theta}{(r-m)^2 - m^2 \cos^2 \theta}
\left [ \left ( r - m \right ) \sin \theta \frac{\partial {U^{(1)}}}{\partial r} 
+ \cos \theta \frac{\partial {U^{(1)}}}{\partial \theta} \right ],\\
\frac{\partial {k^{(1)}}}{\partial \theta}&=&
\frac{2 m \sin \theta}{(r-m)^2 - m^2 \cos^2 \theta}
\left [ - r \left ( r - 2 m \right ) \cos \theta \frac{\partial {U^{(1)}}}{\partial r} 
+ \left ( r - m \right ) \sin \theta \frac{\partial {U^{(1)}}}{\partial \theta} 
\right ].
\end{eqnarray*}

\subsection{Background matching hypersurface}
\label{intback}
The general spherically symmetric embedding  (\ref{eq:gen_sph_supo})
for $\supo^-$ is given explicitly by
\[
\supo^- :\{ t= t_0(\lambda), r =r_0(\lambda), \theta = \vartheta
, \phi = \varphi \},
\] 
where $t_0(\lambda)$ and $r_0(\lambda)$ are smooth functions 
($C^3$ at least) restricted only to the condition that
$\supo^-$ is timelike
(this implies an upper bound for $| \frac{dr_0}{dt_0} |$).
The  coordinate tangent vectors (\ref{eq:gen_es})
to $\supo^-$ read now
\[
\vec e^{~-}_1= \dot t_0\left.\partial_t
+\dot r_0\partial_r\right|_{\supo^-},~~~~~
\vec e^{~-}_2=\left.\partial_\theta\right |_{\supo^-},~~~~~
\vec e^{~-}_3=\left.\partial_\phi\right |_{\supo^-},
\]
and the induced metric on $\supo^-$ is
\[
\fffback{}_{ij}^- d\xi^i d \xi^j =
-N^2_-d \lambda^2
+ r_0^2(\lambda) \left ( d \vartheta^2 + \sin^2 \vartheta d \varphi^2 \right ),
\]
with
\begin{equation}
\label{N}
N^2_- = \left(1-\frac{2m}{r_0}\right)^{-1}
\left[\dot{t}_0^2 \left (1 - \frac{2m}{r_0} \right )^2
- \dot{r}^2_0\right].
\end{equation}
The sign of $N_-(\lambda)$ will be left free for the moment.
The unit normal (\ref{eq:gen_n}) to $\supo^-$
\[
\vnback_{-}=\left.\frac{1}{N_-}\left[\left(1-\frac{2m}{r_0}\right)\dot t_0\partial_{r}
+\left(1-\frac{2m}{r_0}\right)^{-1} \dot
r_0\partial_t\right]\right|_{\supo^-},~~~~~
\fnback_-=\frac{1}{N_-}(-\dot r_0dt+\dot t_0 dr)|_{\supo^-}
\]
points outwards from the interior
Schwarzschild region (increasing $r$) whenever
$\dot t_0>0$ and $N_->0$.
The extrinsic curvature (\ref{eq:sffback}) relative to this normal reads
\begin{eqnarray*}
\sffback{}_{ij}^- d \xi^i d \xi^j = 
\frac{1}{N_-} \left [ \left ( - \dot{t}_0 \ddot{r}_0 + \ddot{t}_0 \dot{r}_0 + \frac{3 m \dot{r}_0^2 \dot{t}_0
}{r_0 \left (r_0- 2 m \right )}
- \frac{m}{r_0^2} \left ( 1- \frac{2m}{r_0} \right ) \dot{t}_0^3 \right )  d \lambda^2  \right . \\
\left . +\dot{t}_0 \left ( r_0 - 2 m \right )
\left ( d \vartheta^2 + \sin^2 \vartheta d \varphi^2 \right ) \right ],
\end{eqnarray*}
which, after comparison with (\ref{eq:sffback}), gives $\K_-$ and $\bar\K_-$ explicitly as
\begin{eqnarray}
  &&\K_- = 
  \frac{1}{N^3_-}
  \left ( - \dot{t}_0 \ddot{r}_0 + \ddot{t}_0 \dot{r}_0
    + \frac{3 m \dot{r}_0^2 \dot{t}_0}{r_0 \left (r_0- 2 m \right )}
    - \frac{m}{r_0^2} \left ( 1- \frac{2m}{r_0} \right ) \dot{t}_0^3 \right ),
  \label{eq:Kin}
  \\
  &&\bar \K_-=\frac{1}{r_0^2 N_-}\dot{t}_0 \left ( r_0 - 2 m \right ).
  \label{eq:barKin}
\end{eqnarray}

\subsection{First order perturbation of the matching hypersurface}
\label{intscalars}
We now derive the perturbed first and second fundamental forms on
$\supo^-$ in terms of scalar quantities. 
We start by considering a general vector 
\begin{equation}
\label{intZ}
\vec{Z}^- = \left.
Z^0 (\lambda,\vartheta,\varphi) \partial_t + 
Z^1 (\lambda,\vartheta,\varphi) \partial_r + 
Z^2 (\lambda,\vartheta,\varphi) \partial_{\theta} + 
Z^3 (\lambda,\vartheta,\varphi) \partial_{\phi}\right|_{\supo^-},
\end{equation}
which describes how the matching hypersurface
$\supo^-$ is deformed to first order.
Using (\ref{pertfirst}) and (\ref{pertsecond}) 
the perturbed first and second fundamental forms on $\supo^-$
can be readily computed. The
results are shown in Appendix \ref{appendixA}.
In order to write down the matching conditions in terms
of $S^2$ scalars, we need to decompose
the vector
$Z^2 \partial_{\vartheta} + Z^3 \partial_{\varphi}$ (which is
tangent to the spherical orbits) according to its Hodge
decomposition. Explicitly
\[
Z^2 \partial_{\vartheta} + Z^3 \partial_{\varphi} =
\left ( \frac{\partial \tinone}{\partial\vartheta}
-\frac{1}{\sin \vartheta} \frac{\partial \tintwo}{\partial\varphi} \right )
\partial_{\vartheta} + \left ( \frac{1}{\sin^2 \vartheta}
\frac{\partial \tinone}{\partial\varphi} 
+ \frac{1}{\sin \vartheta}  \frac{\partial \tintwo}{\partial\vartheta}
\right ) \partial_{\varphi},
\]
where  $\tinone(\lambda,\vartheta, \varphi)$ and 
$\tintwo(\lambda,\vartheta, \varphi)$ are $S^2$ scalars which are defined up to additive 
functions of $\lambda$.
The radial part of $\vec{Z}^-$
can be also decomposed in the following intrinsic manner
\begin{equation}
\label{eq:renamez0z1}
\left. Z^0 \partial_t + Z^1 \partial_r \right|_{\supo}
= \cqin \vnback_-+ \ttauin \vec e_1^-,
\end{equation}
where, again, $\cqin(\lambda,\vartheta,\varphi)$
and $\ttauin(\lambda,\vartheta,\varphi)$ are scalars on $\supo^-$.

When studying the Hodge decomposition of the first order perturbation tensor
$\gfpert{}^-$, we found it convenient to define two
new scalars $\cgsup(r,\theta)$ and $\cpsup(r,\theta)$ by
\begin{eqnarray*}
  A^{(1)} &=&
  \sin \theta \frac{\partial \cgsup}{\partial \theta}, \\
  k^{(1)} &=&\frac{\partial^2 \cpsup}{\partial \theta^2}
    - \frac{\cos \theta}{\sin \theta} \frac{\partial \cpsup}{\partial \theta}
    \equiv \opthp.
\end{eqnarray*}
The function $\cgsup$ is defined up to an additive function of $r$ while the
kernel freedom in $\cpsup$ corresponds to $P_0(r) + P_1 (r) \cos \theta$, for
arbitrary $P_0$ and $P_1$.

As a side remark, we recall that a useful function in any stationary and vacuum spacetime
is the twist potential $\Omega$,
which in the axially symmetric case can be written as
$d A=\rho e^{-2U}*d \Omega$ (where the Hodge dual operator $*$ refers
to the $\{\rho,z\}$ plane). The background value of $\Omega$ is
obviously zero. 
The first order perturbed twist potential $\Omega^{(1)}$ 
is, in fact, closely related to 
the function $\cgsup$ above. 
By appropriately restricting the additive function
in $\cgsup$, it can be checked that
$\Omega^{(1)}=\left(1-\frac{2m}{r}\right) \cgsup_{,r}
$ holds (the freedom left in $\cgsup$ is the addition of
any function $g(r)$ solving
$\left(1-\frac{2m}{r}\right) g_{,r}  = c$ for an 
arbitrary constant $c$).

With all the above definitions, the 
Hodge decomposition of 
the angular components of $\ffffpert{}^{-}_{ij}$ and
$\sfffpert{}^{-}_{ij}$ can be computed. 
The corresponding scalars,
following the notation in (\ref{DefsS2Scalars}), are:
\begin{eqnarray}
\label{int-scalars}
\Fqin &=& r_0^2 \dot \tinone - N_{-}^2 \ttauin , \nonumber\\
\Gqin &=& r_0^2 \dot \tintwo - \dot t_0 \cgsup|_{\supo} \left(1-\frac{2m}{r_0}\right)\nonumber\\
\Pqin &=& r_0^2 (\tinone + \cpsup)|_{\supo}, \nonumber\\
\Rqin &=& r_0^2 \tintwo, \nonumber\\
\Hqin &=& 
\left.2 r_0\left(\frac{\cqin}{N_{-}}\left ( 1- \frac{2m}{r_0} \right )\dot t_0
+\ttauin \dot r_0\right)
- 2 r_0^2 \uperonepsup - 2 r_0^2 \frac{\cos \theta}{\sin \theta}
\frac{\partial \cpsup}{\partial \theta}\right|_{\supo}\nonumber\\
\Gextin &=& \left.\frac{1}{N_{-}}
\left [  \left ( 1-\frac{2m}{r_0} \right )
\left(r_0 \dot t_0 \dot \tintwo - \frac{N_{-}^2}{2 } 
\frac{\partial \cgsup}{\partial r} \right)
- \cgsup \left(\frac{\dot{r}_0^2 r_0+m N_{-}^2}{r_0^2} \right)\right ] \right|_{\supo}\\
\Fextin
&=&\frac{1}{N_{-}}
\left\{
\left ( 1- \frac{2m}{r_0} \right )r_0 \dot t_0\dot \tinone
+\ttauin
\left[
\ddot t_0\dot r_0-\dot t_0 \ddot r_0-\dot t_0 N_{-}^2\frac{m}{r_0^2}
+\dot t_0\dot r_0^2\frac{2m}{r_0(r_0-2m)}
\right]
\right.\nonumber\\
&&\left.
+\frac{\cqin}{N_{-}}
\left[
\dot r_0 \ddot r_0\frac{r_0}{r_0-2m} - \dot t_0 \ddot t_0\frac{r_0-2m}{r_0}
-\frac{2m \dot r_0^3  }{(2m-r_0)^2}
+\frac{N_{-}^2\dot r_0(3m-r_0)}{r_0(2m-r_0)}
\right]
\right.\nonumber\\
&&
\left.\left.
-N_{-}^2\ddl\left(\frac{\cqin}{N_{-}}\right)
+2 \dot r_0 \dot t_0 \uperonepsup
- \dot r_0 \dot t_0 \opthp
\right\}\right|_{\supo}
\nonumber\\
\Pextin &=& \left.\frac{1}{2N_{-}}
  \left [ 
    \left ( r_0 - 2 m \right ) \dot t_0
    \left ( \tinone + \cpsup + r_0 \frac{\partial \cpsup}{\partial r} \right ) - N_{-}\cqin
\right ]\right|_{\supo}, \nonumber\\
\Rextin &=& \left.\frac{1}{N_{-}} \left ( \left ( r_0  -2 m \right )\dot t_0 \tintwo  
+ \frac{\dot{r}_0 }{2}\cgsup
\right )\right|_{\supo}, \nonumber\\
\Hextin
&=&
\frac{1}{N_{-}}\left\{
\ttauin r_0\dot r_0\frac{1}{N_{-}^2}
\left[
  \ddot t_0\dot r_0-\dot t_0 \ddot r_0 -\dot t_0 N_{-}^2\frac{(m-r_0)}{r_0^2}-
  \dot t_0\dot r_0^2 \frac{2m}{r_0(r_0-2m)}
\right]
\right.
\nonumber\\
&&
-\cqin\frac{\dot r_0 r_0}{N_{-}^3}
\left(
\dot r_0 \ddot r_0\frac{r_0}{r_0-2m} - \dot t_0 \ddot t_0\frac{r_0-2m}{r_0}
-
\frac{2m \dot r_0^3  }{(2m-r_0)^2}+\frac{N_{-}^2\dot r_0(m-r_0)}{r_0(2m-r_0)}
\right)
\nonumber\\
&&
+\cqin N_{-} \frac{r_0-m}{r_0}+\dot r_0 r_0\ddl\left(\frac{\cqin}{N_{-}}\right)
+ \dot t_0 \dot r_0^2 r_0 N_{-}^{-2} \left ( \opthp
- 2 \uperonepsup \right )
\nonumber\\
&&
\left.\left. 
+\dot t_0
(2m-r_0) \left ( \DeltaS \cpsup + r_0
\frac{\cos \theta}{\sin \theta}
\frac{\partial^2 \cpsup}{\partial r\partial\theta} + 
\uperonepsup + \deruperonepsupr 
\right )
\right\} \right|_{\supo},\nonumber
\nonumber
\end{eqnarray}
where $\DeltaS$ denotes the Laplacian on $(S^2,h_{AB})$.

\section{Perturbed FLRW region $(+)$}
\label{exterior}
In this section we describe the perturbations of the FLRW region
(denoted by a ($+$) sign)
and derive  the perturbed first and second fundamental forms on the
matching hypersurface $\supo^+$. 
\subsection{First order perturbations of FLRW}
\label{extdecomp}
On a background FLRW spacetime there exists a coordinate system
$\{\tau,x^a\}$ in which the metric reads
\[
\gback{}^{+}=a^2(\tau)\left (-d\tau^2+ \gamma_{ab}dx^a dx^b \right ),
\]
where $\gamma_{ab}dx^a dx^b =
d\rr^2 + \sig^2 (\rr,\curv) ( d \theta^2 + \sin^2 \theta d \phi^2)$
with $\sig=\sinh \rr,\rr,\sin \rr$, for $\curv=-1,0,1$ respectively.
The covariant derivative associated to $\gamma_{ab}$ will be
denoted by $\nabla_a$.

Let us now decompose the first order perturbation tensor $\gfpert{}^+$
into scalar, vector and tensor perturbations
\cite{Stewart}
\begin{eqnarray*}
\gfpert{}^+_{00}&=&-2a^2\Psi\nonumber\\
\gfpert{}^+_{0a}&=&a^2W_a\nonumber\\
\gfpert{}^+_{ab}&=&a^2(-2\Phi \gamma_{ab}+\chi_{ab})
\end{eqnarray*}
with
\[
\chi_{ab}=D_{ab}\cchi+2\nabla_{(a}\cy_{b)}+\ppi_{ab},
\]
where 
\[
D_{ab}\equiv\nabla_{a}\nabla_{b}-\frac{1}{3}\gamma_{ab}\nabla^2,
\]
and
\begin{equation}
  \label{constraint1}
  \nabla^a\cy_a=\ppi_{a}^{~a}=0, \quad \nabla^a\ppi_{ab}=0.
\end{equation}
The vector term $W_a$ can be decomposed further into its irreducible parts:
\[
W_a=\partial_a \wzero+\tilde W_a
\]
with 
\begin{equation}
\label{constraint2}
\nabla^a\tilde W_a=0.
\end{equation} 
The evolution and constraint equations for each mode, in any gauge,
are given e.g. in \cite{NohHwang} 
and can be written in a closed form
after a gauge is specified. In this paper we intend to derive
perturbed matching conditions which can be used in any $(+)$
spacetime gauge. A gauge will only be specified in Appendix
\ref{sec:poisson_gauge} where the Poisson gauge
will be chosen as an example of how
the perturbed matching equations simplify from their general
expressions to a specific gauge.

Next, we introduce $S^2$ scalars 
$\wone$, $\wtwo$, $\yone$, $\ytwo$, $\qone$, $\qtwo$,
$\uone$, $\utwo$ and $\ch$ 
according to 
\begin{eqnarray*}
\tilde{W}_{\theta} d \theta + \tilde{W}_{\phi} d \phi & = &
d \wone + \star d \wtwo, \\
Y_{\theta} d \theta + Y_{\phi} d \phi  & = & 
d \yone + \star d \ytwo, \\
\ppi_{R \theta} d \theta + \ppi_{R \phi } d \phi  & = &
d \qone + \star d \qtwo, \\
\ppi_{AB} 
& = & 
D_A \left ( D_B \uone + (\star d \utwo)_B \right )
+  D_B \left ( D_A \uone + (\star d \utwo)_A \right )
+ \ch h_{AB},
\end{eqnarray*}
where $h_{AB}$, $\star$ and $D_A$ refer here
to the coordinates $\{\theta, \phi\}$.
The trace-free condition on $\ppi_{ab}$ gives 
\[
\ppi_{\rr \rr} =  -\frac{2}{\sig^2} \left (
\DeltaS \uone +  \ch \right ).
\]
As before, these scalars are defined up the kernel of the Hodge
operator, which in this case involves functions of 
$\tau$ and $r$. Concretely, each one of
$\wone$, $\wtwo$, $\yone$, $\ytwo$, $\qone$, $\qtwo$,
admits the freedom
$\wone\to \wone+ w_1(\tau,\rr)$, etc...,
while  $\uone$  (and similarly $\utwo$) is defined up to
$\uone \to \uone + u_1(\tau,\rr)+ u_{1m}(\tau,\rr) Y^m_1$
which implies
$\ch\to \ch+2u_{1m}(\tau,\rr) Y^m_1 $.
Another interpretation of this freedom
is that $\wone$,...,$\qtwo$  do not contribute
to the $l=0$ harmonic sector of the perturbations,
and that $\uone, \utwo, \ch$ do not contribute
to the $l=0,1$ sectors, since one can always
choose the kernels such that $\wone_{(l=0)}=0$, etc...
This is made explicit in the relationship
between these scalar functions and the harmonic decompositions
in \cite{Mukohyama,MMV1}, as shown in Appendix \ref{variable-identif}.

The constraints (\ref{constraint1}) and (\ref{constraint2})
in terms of the Hodge scalars read
\begin{eqnarray*}
\frac{1}{\sig^2} \DeltaS  {\yone} 
+ \frac{2 \dersig}{\sig} \ychi
+ \derxiychi & = & 0, \label{equationyy} \\
\frac{1}{\sig^2} \DeltaS  {\wone} 
+ \frac{2 \dersig}{\sig} \wchi 
+  \derxiwchi & = & 0, \label{equationww} \\
\DeltaS \qone - 2 \DeltaS \derxiuone- \frac{2 \dersig}{\sig} 
\left ( \DeltaS  \uone  +  \ch \right ) - 2  \derxich & = & 0,
\label{equationppichi}  \\
2 \DeltaS \uone + 2 \uone + \ch + \sig^2 \derxiqone
+ 2 \sig \dersig \qone - \bone(\tau,\rr) & = & 0, \label{equationppi1} \\
\DeltaS \utwo + 2 \utwo +  \sig^2 \derxiqtwo
+ 2 \sig \dersig \qtwo - \btwo(\tau,\rr) & = & 0, \label{equationppi2} 
\end{eqnarray*}
where the prime denotes derivative with respect to $R$ and
$\bone$ and $\btwo$ are arbitrary functions. The latter functions arise
because we need to Hodge decompose the equations $\nabla^a \ppi_{a B}  = 0$,
from which two extra kernel functions appear.

The relationship between the Hodge scalars and the Mukohyama variables for 
perturbations of spherical backgrounds is summarized in
Appendix \ref{variable-identif}.

\subsection{Background matching hypersurface}
\label{extback}
In the Einstein-Straus and Oppenheimer-Snyder models
the matching hypersurface is
comoving with respect to the FLRW flow. In fact, it is now
known that this is necessary for any matching of a static,
or stationary and axisymmetric, vacuum region to a FLRW spacetime
\cite{Mars98,Mars01,Nolan-Vera}. The matching hypersurface
$\supo^+$ is therefore of the form
\[
\supo^+: \{ \tau=\lambda, R = R_c, \theta = \vartheta
, \phi = \varphi \},
\]
where $R_c$ is a constant. The tangent vectors
are
\[
\vec e^+_1= \left.\partial_\tau\right|_{\supo^+},~~~~~
\vec e^+_2=\left.\partial_\theta\right |_{\supo^+},~~~~~
\vec e^+_3=\left.\partial_\phi\right |_{\supo^+},
\]
and the 
first fundamental form is
\[
\fffback{}^+_{ij}d\xi^i d \xi^j =  \asig^2
\left[ 
-d \lambda^2 +
\sigo^2
\left ( d \vartheta^2 + \sin^2 \vartheta d \varphi^2 \right )
\right] ,
\]
where $\sigo \equiv \sig(R_c,\curv)$
and $\asig\equiv a|_{\supo}=a(\lambda)$.
Comparing this expression with (\ref{eq:fffback}) we have
$r^+|_{\supo}=\asig\sigo$ and
$N^2_+=\asig^2$. 
The unit normal to $\supo^+$ pointing towards the
direction in which $R$ increases reads 
\[
\vec n^+= \left . \frac{1}{\asig} \partial_R \right |_{\supo},~~~~~ \bm{n}^+ = \asig dR|_{\supo},
\]
and a simple calculation gives the second fundamental form on $\supo^+$ to be
\[
\sffback{}_{ij}^+ d \xi^i d \xi^j =
 \asig \sigo \dersigo \left ( d \vartheta^2 +
\sin^2 \vartheta^2 d \varphi^2 \right ),
\]
where $\dersigo \equiv \dersig(R,\curv) |_{R =R_c}$.
Comparing this expression with (\ref{eq:sffback})
we find 
\begin{equation}
\K_+=0, \quad \bar\K_+=\dersigo/(\asig\sigo). \label{eq:Kout}
\end{equation}

\subsection{First order perturbation of the matching hypersurface}
\label{extscalars}
The first order 
perturbation of $\supo^+$ is defined by a vector field $\vec{Z}^+$
at points on $\supo^+$. Similarly to the case of the Schwarzschild region, we
decompose $\vec{Z}^+$ as
\[
\vec{Z}^+ = \left.\ttauout \partial_{\tau} + \frac{\cqout}{a} \partial_R
+ \left ( \frac{\partial \tone}{\partial \vartheta} - \frac{1}{\sin \vartheta}
\frac{\partial \ttwo}{\partial \varphi} \right ) \partial_{\theta} 
+ \left ( \frac{1}{\sin^2 \vartheta} \frac{\partial \tone}{\partial \varphi}
+ \frac{1}{\sin \vartheta}
\frac{\partial \ttwo}{\partial \vartheta} \right )
\partial_{\phi}\right|_{\supo^+}, 
\]
where $\ttauout$, $\cqout$, $\tone$ and $\ttwo$ depend on
$\{\lambda,\vartheta,\varphi\}$.
The Hodge decomposition of
the angular parts of $\ffffpert{}^+_{ij}$ and $\sfffpert{}^+_{ij}$
(with explicit expressions given in Appendix \ref{appendixB})
in terms of the $S^2$ scalars introduced above
can be found after a straightforward but somewhat long calculation.
Recalling the notation in (\ref{DefsS2Scalars}), the result
is\footnote{At some points we slightly abuse the notation and use dot
to denote both derivative with respect to $\tau$, and derivative with
respect to $\lambda$. On the matching hypersurface they obviously
coincide as $\tau=\lambda$ and $R,\theta,\phi$ do not
depend on $\lambda$ there.}
\begin{eqnarray}
\label{out-scalars}
\Fqout &=& \left.\asig^2 \left (\dot \tone \sigo^2 - \ttauout
+ \wone + \wzero \right )\right|_{\supo}, \nonumber\\ 
\Gqout &=& \left. \asig^2 \left (
\dot\ttwo \sigo^2 + \wtwo \right )\right|_{\supo}\nonumber\\
\Pqout &=& \left.\asig^2
\left ( \frac{1}{2} \cchi + \sigo^2 \tone + \uone + 
\yone \right )\right|_{\supo},  \nonumber\\
\Rqout &=& \asig^2 \left.(\sigo^2 \ttwo + \utwo + \ytwo)\right|_{\supo}, 
\nonumber\\
\Hqout &=& 
\asig^2 \left ( 
- \frac{1}{3} \DeltaS \cchi
+ \ch 
+ 2   \dersigo \sigo\ychi 
-  \frac{1}{3} \derxixichi \sigo 
- 2 \Phi \sigo^2 
+ \frac{1}{3} \dersigo\sigo  \derxichi \right ) \nonumber\\
& & \left.+ 2 \asig \left ( \cqout \dersigo\sigo
+ \asigdot  \sigo^2 \ttauout  \right )\right|_{\supo}\nonumber\\
\Fextout &=& 
 \frac{\asig}{2} \left (
\derxiwone - \wchi
- \dot \derxichi
- \dot \derxiyone 
- \dot \qone
- \dot \ychi 
\right ) \\
&&\left.
+ \frac{\asig \dersigo}{\sigo} \left (
 \frac{1}{2} \dot \cchi
+ \dot \yone 
+ \sigo^2  \dot {\tone}  
\right ) 
- \dot\cqout
+ \frac{\asigdot}{\asig} \cqout \right|_{\supo} \nonumber\\
\Gextout &=& \left.
\frac{\asig}{2} \left ( 
\derxiwtwo 
- \dot \derxiytwo 
- \dot\qtwo
\right ) + 
\frac{\asig \dersigo}{\sigo} \left (
\dot\ytwo + \sigo^2   \dot\ttwo
\right )\right|_{\supo}\nonumber\\
\Pextout &=& \left.
\frac{\asig \dersigo}{\sigo} \left ( \yone  
+ \frac{1}{2} \cchi  \right )
+ \frac{\asig}{2} 
\left ( 2 \dersigo \sigo \tone  
+ \derxiuone  - \qone  - \ychi
- \frac{1}{2} \derxichi \right )
- \frac{\cqout}{2} \right|_{\supo}\nonumber\\
\Rextout &=& \asig \left.\left (
 \dersigo \sigo \ttwo 
+  \frac{\dersigo}{\sigo} \ytwo 
+ \frac{1}{2}\derxiutwo  - \frac{1}{2} \qtwo \right )\right|_{\supo}
\nonumber\\
\Hextout &=& 
\frac{\asig \dersigo}{\sigo} \left ( \frac{1}{6} \DeltaS \cchi
+ \DeltaS \uone + \ch 
\right )
-  \frac{\asigdot^2}{\asig^2} \sigo^2 \cqout
+  \frac{\asigdot}{\asig} \sigo^2 \dot \cqout 
+  \asigdot \dersigo \sigo \ttauout \nonumber\\
& &
+  \asigdot \sigo^2 \left ( \wchi + \derxiwzero \right ) 
+ \asig \left ( 
\frac{1}{2}   \derxichi 
+ \frac{1}{2}   \derxich  
+  \ychi + \frac{\cqout}{\asig}
- \frac{1}{6}  \DeltaS \derxichi \right . \nonumber\\
& & \left.
\left . -  \dersigo \sigo \left ( \frac{1}{2} \derxixichi + \Phi \right )
- \sigo^2 \left (
 \frac{2 \curv }{3}  \derxichi
+  \derxiphi
+  \frac{1}{6} \derxixixichi 
+ 2 \curv \ychi
+ 2   \curv \frac{\cqout}{\asig}  \right ) \right )\right|_{\supo}.\nonumber
\end{eqnarray}
This concludes the decomposition of the perturbation. Our next aim
is to write down and discuss 
the matching conditions.

\section{Matching conditions}
\label{MainSection}
\subsection{Background matching conditions: the Einstein-Straus
and Oppenheimer-Snyder models}

The results in this subsection are well-known,
but we reproduce their derivation for completeness.
The background matching conditions are obtained
simply by particularising the equations (\ref{eq:backmc2}) (which
correspond to (\ref{eq:backmc}) in spherical symmetry) to the
Schwarzschild region and the FLRW region.  The second equation in
(\ref{eq:backmc2}) implies
\begin{equation}
  r_0 = \sigo \asig
  \label{Back1}.
\end{equation}Inserting this into (\ref{N}), and using its derivative
along $\lambda$, i.e. $\dot r_0=\sigo\dot\asig$, the first equation
in (\ref{eq:backmc2}) leads to a
quadratic equation for $\dot{t}_0$, namely
\begin{equation}
\dot{t}^2_0 = \frac{\sigo^2 \dot{\asig}^2 + \asig^2
- \frac{2m}{\sigo} \asig}{\left (
1 - \frac{2m}{\sigo} \asig \right )^2 }. \label{t02}
\end{equation}
From $ N_{-}^2 = \asig^2$ we write
$N_{-}=\sigma \asig$ with $\sigma = \pm 1$ (recall that 
we want to keep the
orientation of the normal arbitrary).
The fourth equation in  (\ref{eq:backmc2}), together with
expressions  (\ref{eq:barKin}) and (\ref{eq:Kout}) give
the following  linear equation for $\dot{t}_0$
\begin{equation}
\dot{t}_0 = \sigma \frac{\asig^2 \sigo \dersigo}{\sigo \asig - 2 m }
\label{Back2},
\end{equation}
which inserted into (\ref{t02}) yields
\[
\dot{\asig}^2 + \curv \asig^2 = \frac{2m \asig}{\sigo^3},
\]
after using ${\dersigo}^2 = 1- \curv \sigo^2$. This ODE
for the scale factor is exactly the Friedmann equation for dust (restricted
to points on $\supo$), as expected. This is just a consequence
of the Israel conditions,  which impose the equality of
certain components of the 
energy-momentum tensor on both sides of the matching hypersurface.

The last matching condition, namely 
the third equation in (\ref{eq:backmc2}), is automatically
fulfilled once (\ref{Back1}), (\ref{Back2}) and the dust Friedmann
equation hold, as a straightforward calculation shows.

Summarizing, given the necessary condition that the 
FLRW background is dust, the matching conditions are
satisfied if and only if
(\ref{Back1}) and (\ref{Back2}) hold. We can now study the
linearly perturbed matching.

\subsection{Linearised matching}
\label{MainSubSection}
The linearised matching conditions (\ref{eq:fpertmc}) correspond to
equating the expressions for
$\ffffpert{}^\pm_{ij}$ and
$\sfffpert{}^\pm_{ij}$ given in Appendices \ref{appendixA} and
\ref{appendixB}. However, as discussed in
Section \ref{sec:scalar-match}, the angular components
are much better handled if 
the underlying spherical symmetry is exploited
through the Hodge decomposition, which allows us to work exclusively
in terms of $S^2$ scalars. Thus, the full set of matching
conditions is given by
(\ref{eq:match-scalars}) after using (\ref{int-scalars}) and
(\ref{out-scalars}), together with the non-angular expressions
(\ref{eq:q1ll_in}), (\ref{eq:q1ll_out}),
(\ref{eq:k1ll_in}), and (\ref{eq:k1ll_out}),
given in Appendices \ref{appendixA} and \ref{appendixB}.

The Hodge decomposition in terms of scalars involves two types
of objects depending on their behaviour under reflection.
For instance, in the decomposition (\ref{Hodge})
the scalar $F$ remains unchanged while $G$ changes
sign under a reflection. The former scalar is then named {\it
even} while the latter is named {\it odd}. This splitting
behaviour occurs
in any decomposition in terms of Hodge potentials.
In particular, our linearised matching conditions must split into 
equations involving only even scalars and
equations involving only odd scalars. We denote them
simply as the even and odd sets of equations.

The odd set is simpler to handle. It is not difficult
to see that the equations can be rewritten
as the following four relations
\begin{eqnarray}  
\label{eq:ttwo}
\ttwo + \sigo^{-2}[\uytwo] & \eqq &  \tintwo, \\
\label{eq:W2}
\wwtwo  
& \eqq &  -  \sigma \cgsup \dersigo \asig^{-1},\\
\label{eq:dW2}
\dwwtwo 
& &
\nonumber\\
\eqq \sigma \cgsup \frac{\sigo^3 \asig\curv -3m}{\sigo^2 \asig^2}
& +& 
\sigma\dercgsupr(\sigo^2\curv-1),\\
\label{eq:Q2}
\qtwo -[\derxiutwo - {\markkern  2\asig (\kernelRext)}]
+2\sigo^{-1}\dersigo[\utwo- {\markkern  (\kernelRq)}]
&\eqq&  -\sigma\cgsup \asig^{-2}\asigdot \sigo.
\end{eqnarray}
The even set of equations requires a much more lengthy and 
subtle analysis. After carefully combining the equations,
and defining $\difcqcqin\equiv \cqout-\cqin$, it turns
out that they can be written as the following eight equations
\begin{eqnarray}
\label{eq:tone}
&&\tone + \sigo^{-2}[\uyone] \eqq \tinone + \cpsup,\\
\label{eq:ttau}
&&\ttauout  -[\wwone] \eqq \sigo^3 \asig \dercpsupr +\ttauin, \\
\label{eq:Q1}
&&\qone +\ychi +\derxichi
-[{{\cal U}_1}' - {\markkern  2a_{\Sigma} (R^{\kappa}_{0}+R^{\kappa}_{m} Y_1^m)}]\nonumber\\
&&~~~~
+2 \sigo^{-1}\dersigo[{{\cal U}_1} - {\markkern  (P^q_{0}+P^q_{m} Y_1^m)}]+\diffqq \eqq -\dersigo\sigo^2 \asig \dercpsupr,\\
\label{eq:Psi} 
&& \Psi
+\frac{1}{\asig}\ddl\left[\asig\left.\left( \wwone\right)\right|_{\supo}\right]
\nonumber\\
&&~~~~ 
\eqq \frac{\asig\sigo + 2 m - 2 \asig \sigo^3 \curv }{\asig \sigo - 2m}\uperonepsup
+(2\sigo^3 \asig \curv -3m) \dercpsupr
+\sigo\asig(\sigo^3 \asig \curv -2m) \der2cpsupr\nonumber\\
&&~~~~
+ \frac{ \asig \sigo^3 \curv -2 m }{ \asig \sigo - 2 m }\opthp,\\
\label{eq:dW1}
&&\dwwone \nonumber\\
&&~~~~ \eqq
\frac{2\asigdot\sigo^2\dersigo}{\asig\sigo-2m}(2\uperonepsup-\opthp)
-3\asigdot \sigo^2\dersigo\dercpsupr-\asigdot\asig\sigo^3\dersigo\der2cpsupr, \\
\label{eq:Psiprime}
&&\derxipsi+\frac{1}{\asig}
\ddl\left(\asig\left[\wchi+\derxiwzero+\ddl\left(\diffqq\right)\right]|_{\supo}\right)
\eqq
-\frac{3m}{\asig^2\sigo^3}\cqin \nonumber \\
&&~~~~
+ \frac{2\sigo^3\asig\curv+\asig\sigo-6m}{\asig\sigo-2m}
\asig \dersigo \deruperonepsupr
-\frac{\sigo^3\asig\curv-2m}{\asig\sigo-2m}\asig \dersigo
\frac{\partial}{\partial r}\opthp    \nonumber\\
&&~~~~
-2\frac{\sigo^2\curv-1}{(\asig\sigo-2m)^2}\asig\dersigo (2\uperonepsup-\opthp),
\\
\label{eq:Phiprime}
&&\derxiphi+\frac{1}{6}\derxixixichi
+ \frac{\dersigo}{2\sigo}(\derxixichi+2\Phi)
-\frac{\asigdot}{\asig}(\wchi + \derxiwzero) \nonumber\\
&&~~~~
- \frac{\asigdot\dersigo}{\asig\sigo}\left(\wwone\right) \nonumber\\
&&~~~~
-\frac{\dersigo}{\sigo^3}[\ch + \DeltaS\uone
+\frac{1}{6}\DeltaS\cchi]
-\frac{1}{2\sigo^2}[\derxich - {\markkern  4\asig \Pextm Y_1^m}] \nonumber\\
&&~~~~
+\frac{1}{6\sigo^2}(\DeltaS\derxichi-\derxichi(3 -4 \curv \sigo^2))
-\ychi\left(\sigo^{-2}-2\curv\right)
\nonumber\\
&&~~~~\eqq 
\frac{1}{\sigo^2}(1-2\sigo^2\curv) \diffqq
+ \frac{1}{\asigdot \sigo^3}(2m-2\asig\sigo^3\curv) \ddl\left(\diffqq\right)
-\frac{3m}{\asig^2\sigo^3}\cqin
\nonumber\\
&&~~~~
+\frac{\dersigo(\asig\sigo+2m-2\asig\sigo^3\curv)}{\sigo(\asig\sigo - 2m)}
\uperonepsup
+\asig\dersigo \deruperonepsupr
-\frac{\dersigo}{\sigo}(2m-\asig\sigo^3\curv)\dercpsupr\nonumber\\
&&~~~~
+\frac{\dersigo(\asig\sigo+\asig\sigo^3\curv-4m)}{\sigo(\asig\sigo-2m)}\opthp
+\frac{\dersigo}{\sigo}\frac{\cos\theta}{\sin\theta}
\frac{\partial}{\partial\theta}\left(2\cpsup+\asig\sigo\dercpsupr\right),\\
\label{eq:diffqq}
&&\diffqq+\frac{\asigdot\sigo}{\asig\dersigo}\left(\wwone\right)\nonumber\\
&&~~~~
-\frac{1}{6\sigo\dersigo}\DeltaS\cchi
+\frac{1}{2\sigo\dersigo}(\ch- {\markkern  2\Pqm Y_1^m})
+\frac{1}{6}\derxichi
+\ychi
-\frac{\sigo}{\dersigo}\left(\frac{\derxixichi}{6}+\Phi\right)\nonumber\\
&&~~~~
\eqq\frac{\sigo}{\dersigo}\left((\sigo^3\asig \curv-2m)\dercpsupr
-\frac{\cos\theta}{\sin\theta}\frac{\partial \cpsup}{\partial \theta}-\uperonepsup\right).
\end{eqnarray}
This set of twelve equations
represent the full set of linearised matching conditions
for our problem. They are valid for any 
FLRW gauge and any hypersurface gauge. Moreover, they include the
twenty kernel functions 
${\markkern \Fqo}$, ${\markkern \Gqo}$, ${\markkern \Pqo}$,
${\markkern \Rqo}$, ${\markkern \Pqm}$, ${\markkern \Rqm}$,
${\markkern \Fexto}$, ${\markkern \Gexto}$,
${\markkern \Pexto}$, ${\markkern \Rexto}$,
${\markkern \Pextm}$, ${\markkern \Rextm}$
in order to allow for any choice of Hodge decomposition on either
($\pm$) side. Depending on the problem, these kernel functions
may play a role. For instance, if the aim is to determine
perturbations in FLRW given perturbations in the Schwarzschild
region, then the kernel functions can be put to zero without loss of generality
since, in that problem, one is constructing the exterior data and changing the kernel
functions does not affect the metric perturbations. However, in a
situation when two specific perturbations are given and the problem
is to determine whether they match at the linear level, then  the
kernel functions become relevant and cannot be dropped a priori.

The expressions above are written in such a way that
the $(+)$ and the
$(-)$ objects
are kept on the left
and right hand sides of the equations, respectively. The only
exception being the difference $\difcqcqin$ which we found convenient
to use with a pivotal role.

The three equations (\ref{eq:ttwo}), (\ref{eq:tone}), (\ref{eq:ttau})
determine the difference vector $\vec T^+ -\vec T^-$. Recall that this
difference vector is tangent to the background matching hypersurface
and corresponds to the freedom in perturbing points within the
hypersurface without deforming it as a set of points
\cite{MMV1}. Recall also that by choosing the appropriate hypersurface
gauge one can fix either $\vec T^+$ or $\vec T^-$ arbitrarily (but not
both), and that the difference $\vec T^+ -\vec T^-$ is independent of
such choice.
Thus, the three equations (\ref{eq:ttwo}), (\ref{eq:tone}),
(\ref{eq:ttau}) do not provide any essential information concerning
the metric perturbations at either $(\pm)$ side or the shape of the
perturbed $\supo$ (defined by $\difcqcqin$).  We have been careful in
rearranging the remaining equations so that the difference $\vec T^+
-\vec T^-$ does not appear. So, this somewhat superfluous information
gets, in this way, separated from the remaining (more relevant)
restrictions.

We summarize the results of this section in the form of a theorem
\begin{theorem}
\label{MainTheorem}
Let an Einstein-Straus or Oppenheimer-Snyder spacetime geometry be
linearly perturbed in such a way that
the perturbations inside the Schwarzschild region are 
stationary, axially symmetric and vacuum, while the
perturbations of the matching hypersurface and of the FLRW region
are arbitrary. Assume also that the Weyl gauge has been chosen for
the Schwarzschild perturbation and that the
Hodge decomposition has been used to write all tensors on the sphere in 
terms of scalars.

Then, the linearised matching conditions are satisfied (and hence a
perturbed model is obtained) if and only if the equations
(\ref{eq:W2}), (\ref{eq:dW2}), (\ref{eq:Q2}) for the odd part, and the
equations (\ref{eq:Q1}), (\ref{eq:Psi}), (\ref{eq:dW1}),
(\ref{eq:Psiprime}), (\ref{eq:Phiprime}), (\ref{eq:diffqq}) for the
even part are fulfilled.
\end{theorem} When a specific gauge is used on the FLRW side, the
equations above simplify (sometimes notably).  As an example, we
present in Appendix \ref{sec:poisson_gauge} the linearised matching
conditions for the particular case of a flat $\curv=0$ FLRW region in
the Poisson gauge, for which $\wzero=\cchi=\cy_a=0$.

It is also worth noticing that the Einstein equations have been used
at the background level, but the linearised equations for dust at the
FLRW region have not been used anywhere.  Thus, the equations apply to
any perturbation of FLRW \textit{regardless of the matter content
being described}.  The same comment applies to the Schwarzschild side
except for the fact that the perturbations have been restricted a
priori to being stationary and axially symmetric and that the form
(\ref{Weylform}) uses part of the vacuum field equations (in
particular, it uses the fact that $\rho$ is a flat harmonic function,
which allows the metric to be written in the form (\ref{staticWeyl})).
The reader may have also noticed that while the gauge of the FLRW is
kept free, the gauge in the Schwarzschild cavity has been fixed from
the very beginning. The reason for such a different treatment is that
we implicitly regard the perturbed Schwarzschild metric as a {\it
source} for the FLRW perturbations, which then become the unknowns.

Having obtained the equations in a simple and compact form (they may
be compared with the equations that would result from 
equating all components in 
$\ffffpert{}^{-}_{ij}$ and $\sfffpert{}^{-}_{ij}$ in Appendix
\ref{appendixA} 
with their pairings 
$\ffffpert{}^{+}_{ij}$ and $\sfffpert{}^{+}_{ij}$ in Appendix
\ref{appendixB}), our aim now is to extract some of their direct
consequences. A more detailed analysis of the equations is postponed
to a subsequent paper.

\section{Constraints on the FLRW side}
\label{sec:constraints}

The first important general consequence of the linearised
matching equations
arises by simply considering equations (\ref{eq:W2}) and (\ref{eq:Q2}).
Isolating $\cgsup$ from both equations we arrive at the relation
\begin{eqnarray}
  &&  \frac{\asigdot\sigo}{\asig\dersigo}\left(\wwtwo\right)\nonumber\\
  && ~~~~~~~\eqq \qtwo 
  -[\derxiutwo - {\markkern  2\asig (\kernelRext)}]
+2\sigo^{-1}\dersigo[\utwo- {\markkern  (\kernelRq)}],
  \label{eq:constraint1}
\end{eqnarray}
which only involves objects on the FLRW side. Therefore,
this equation constitutes a constraint on the FLRW perturbations on $\supo$,
irrespective of the (stationary and axisymmetric) perturbations
of the Schwarzschild region, and links the vector perturbations
represented by the gauge invariant vector perturbation\footnote{Note
that $W_a-d Y_a/d\tau$ is the (only) gauge invariant vector linear perturbation
\cite{Stewart}.
$\wtwo-d\ytwo/d\tau $ corresponds, then, to the divergence-free and odd part
of the  vector perturbation in FLRW.}
$\wtwo-d\ytwo/d\tau $ 
and the tensor perturbations driven by $\utwo$ and $\qtwo$.

Note also that although
there are kernel terms in the form of ${\markkern\kernelRq}$ and
${\markkern\kernelRext}$,
which would contribute to the $l=0,1$ harmonics, these can be
in principle absorbed into $\utwo$ and $\derxiutwo$ respectively
and do not affect the value of the tensor perturbations.
Thus, equation (\ref{eq:constraint1}) implies that if
there are no tensor perturbations, i.e. $\utwo=\qtwo=0$
then, on $\supo$, $\wtwo-d\ytwo/d\tau $ cannot contain harmonics with $l\geq 2$.

In terms of the doubly gauge invariant
perturbation variables of Mukohyama \cite{Mukohyama,MMV1},
defined for $l\ge 2$,
the constraint
(\ref{eq:constraint1}) restricted to $l\ge 2$ is equivalent to the set
of equations
\begin{equation}
\label{eq:constr_muko}
\frac{\asigdot\sigo}{\asig^2\dersigo}
f^+_0  \eqq -2
\kappa^+_{(LT)}\hspace{1cm} \mbox{ for all ($l\geq 2$, $m$), }
\end{equation}
as it can be easily checked by using the
relations in Appendix \ref{variable-identif} together with (\ref{out-scalars}),
and the expressions for the doubly gauge invariants in \cite{Mukohyama}.

The main result of this section is then summarised in the following theorem:
\begin{theorem}
\label{res:theorem_constraint}
Let a region of a general perturbed dust FLRW be (perturbatively) matched
across a non-null hypersurface to a region
of a stationary and axisymmetric, vacuum perturbation of Schwarzschild.
If the perturbed FLRW contains a vector perturbation with $l\geq 2$
harmonics on $\supo$,
then the FLRW region must also contain tensor perturbations on $\supo$.
\end{theorem}
In other words, \emph{if the FLRW side contains rotational perturbations, then
it must also contain gravitational waves, irrespective of the matter
content described by the perturbation}.

At points where $\asigdot\neq 0$
there is a second constraint on the FLRW side which is obtained
by differentiating (\ref{eq:W2}) along $\lambda$ and using
(\ref{eq:dW2}) to isolate
$\cgsup_{,r}|_{\supo}=\ddl(\cgsup|_{\supo})/(\asigdot\sigo)$.
This second constraint relates the values on $\supo$ of
$\wtwo-d\ytwo/d\tau$ and $\utwo$,
with their first and second derivatives, and reads
\begin{eqnarray}
  \label{eq:constraint2}
  &&
  \asigdot\sigo
  \left\{
    (1-\asig\dersigo)
    \left[\derxiwtwo-\ddl(\derxiutwo+\derxiytwo|_{\supo})\right]
    {\markkern  - 2\asig\Gexto +}
    {\markkern  \ddl[2\asig(\kernelRext)]}\right.\nonumber\\
  &&\left.
    -\left(\wwtwo\right)(3m+\asig\sigo-2\curv\asig\sigo^3)
    \frac{\asigdot}{\asig\sigo\dersigo}
  \right\}\nonumber\\
  &&
  -\asig\dersigo
  \left\{
    \dot \wtwo-\ddl(\dot\utwo+\dot\ytwo|_{\supo})
    -(\derxiutwo+\derxiytwo)\frac{m-\curv\asig\sigo^3}{\sigo^2}
    {\markkern  -\dot\Gqo+\ddot\Rqo+\ddot\Rqm Y_1^m}
  \right\}\eqq 0.
\end{eqnarray}

\section{Matching perturbed Schwarzschild with exact FLRW}
\label{ExactFLRWPertSch}
Nolan and Vera \cite{Nolan-Vera} have proved that in order to match
stationary axially symmetric vacuum regions to FLRW regions across
hypersurfaces preserving the axial symmetry, the vacuum part must be
static. As an application of our formalism above, we shall
show that their result can be generalised to arbitrary matching hypersurfaces
(not necessarily axially symmetric) to first order in approximation
theory.

Since we want to keep the FLRW exact, 
we set all the FLRW perturbations equal to zero. Notice that 
this entails a choice of Hodge scalar functions in the FLRW part,
and therefore
the kernel functions in the matching conditions must be kept free.

Let us start by
considering equation (\ref{eq:W2}),
which differentiated with respect to $\vartheta$
and using the fact that $\cgsup$ does not depend on $\phi$,
leads to ${\markkern  \dot\Rqtwo}
= 
{\markkern  \dot\Rqthree} =0$ plus
\begin{equation}
  \label{eq:a1pre}
  A^{(1)}|_{\supo}= -\sigma \frac{\asig\sin^2{\vartheta}}{\dersigo} 
{\markkern \dot \Rqone}.
\end{equation}
In order to determine ${\markkern  \dot \Rqone}$ we use
the constraints
(\ref{eq:constraint1}) and (\ref{eq:constraint2})
derived in the previous section. Setting all
FLRW perturbations equal to zero, and extracting
the coefficient in $Y^1_1$ of equation 
(\ref{eq:constraint1}) we get
 \[
\Rextone =\frac{\asigdot \sigo}{2\asig^2\dersigo} {\markkern \dot\Rqone}
+\frac{\dersigo}{\asig \sigo} {\markkern \dot\Rqone}.
\]
Inserting this expression into the 
$Y^1_1$ coefficient of the
second constraint (\ref{eq:constraint2}) yields
a second order ODE for ${\markkern \Rqone(\lambda)}$, namely
\[
\asig(2m-\asig\sigo){\markkern \ddot \Rqone}+\asigdot (\asig \sigo-4m){\markkern \dot \Rqone} =0,
\]
which can be solved to give
\begin{equation}
{\markkern \dot\Rqone} = \frac{ C \asig^2}{2 m - \asig\sigo},
\label{GeneralSolution}
\end{equation}
where $C$ is an arbitrary integration constant. Thus, 
(\ref{eq:a1pre}) becomes
\begin{equation}
  A^{(1)}|_{\supo}=
  \sigma C \frac{\asig^3\sin^2{\vartheta}}{\dersigo(\asig\sigo-2m)}.
  \label{ExactFLRW}
\end{equation}
Our aim is to show that the interior source must be static in this case. Since we are working at the
perturbative level, first of all we need to determine the necessary
and sufficient condition that ensures that a given
perturbation of a static background remains static to that
order of approximation. To do that, consider an arbitrary metric
$g_{\alpha\beta}$ with a static Killing vector $\vec{\xi}$ and
a first order perturbation metric
$g^{(1)}_{\alpha\beta}$ which admits a perturbative Killing vector,
i.e. there exists a vector $\vec{\xi}^{(1)}$ such that
$\vec{\xi} + \vec{\xi}^{(1)}$ is (to first
order) a Killing vector of $g_{\alpha\beta} + 
g^{(1)}_{\alpha\beta}$.  Our aim is to find the condition that
has to be imposed to ensure that this vector is static
(to first order). 

Let us, first of all, lower its indices and define
$\hat{\xi}_{\alpha} = (g_{\alpha\beta} +  g^{(1)}_{\alpha\beta} )
(\xi^{\beta} +  \xi^{(1) \, \beta}) = \xi_{\alpha} + 
 g^{(1)}_{\alpha\beta} \xi^{\beta} +  \xi^{(1)}_{\alpha} +
O (2)$. The perturbed Killing
vector is static (to first order) if and only if
the linear term of $\hat{\xi}_{[
\alpha} \partial_{\beta} \hat{\xi}_{\gamma ] }$ vanishes, or
equivalently 
%
%
\[
\bm{M} \wedge d \bm{\xi} + \bm{\xi} \wedge d \bm{M} = 0,
\]
where
$\bm{M} = \bm{\xi^{(1)}} + \bm g^{(1)}_t$, and
$g^{(1)}_t{}_\alpha\equiv \xi^\beta g^{(1)}_{\beta\alpha}$. Now, 
staticity of $\bm{\xi}$ is equivalent to $\bm\xi\wedge\bm w=d\bm\xi$
for some one-form $\bm w$, which allows us to rewrite the
linear staticity condition as
\begin{equation}
  \label{eq:static_1st}
  \bm\xi\wedge(d\bm M - \bm M \wedge\bm w)=0.
\end{equation}
By construction, this equation gives the necessary and sufficient condition
for having a static first order perturbation.

If the static background is moreover axially symmetric and the 
perturbation is stationary and axially symmetric, with no further symmetries,
the vector $\vec{\xi}^{(1)}$ is restricted to having the form 
$\vec{\xi}^{(1)} = a \vec{\xi} + b \vec{\eta}$, where 
and $a$ and $b$ are arbitrary constants, and $\vec{\eta}$ is the axial
Killing vector of the background 
that remains preserved in the perturbation (i.e. the one
fulfilling $\pounds_{\vec{\eta}} g^{(1)}_{\alpha\beta} = 0$).

For a static background of the form (\ref{staticWeyl}) (with $A=0$) we have
\[
\bm\xi=N_{\xi}d t,~~~
\bm\eta=N_{\eta}d \phi,
\]
where $N_\xi$ and $N_\eta$ are respectively the norms
of $\vec\xi$ and $\vec\eta$. For a Schwarzschild background they
are given by
\begin{equation}
\label{eq:Ns}
N_\xi=-\left(1-\frac{2m}{r}\right),~~~
N_\eta=r^2\sin^2\theta\left(1-\frac{2m}{r}\right)^{-1}.
\end{equation}
Computing $d\bm\xi=N_\xi^{-1} d N_\xi\wedge\bm\xi$ we obtain
$
\bm w=-N^{-1}_\xi d N_\xi.
$
On the other hand, using (\ref{Weylform}) for the metric perturbation, we 
find
\[
\bm g^{(1)}_t=A^{(1)}\frac{N_\xi}{N_\eta} \bm \eta + 2 U^{(1)} \bm \xi,
\]
and thus
\[
\bm M=\left(a+2 U^{(1)}\right)\bm\xi+
\left(b+A^{(1)}\frac{N_\xi}{N_\eta}\right)\bm \eta.
\]
It is now straightforward to rewrite 
(\ref{eq:static_1st}) as
\[
d \left(A^{(1)}\frac{N_\xi}{N_\eta}\right)+
\left(b+A^{(1)}\frac{N_\xi}{N_\eta}\right)
\left[\frac{1}{N_\eta} d N_\eta-\frac{1}{N_\xi} d N_\xi\right]=0,
\]
which simplifies to
\[
d A^{(1)}=b \frac{N^2_\eta}{N^2_\xi} d\left(\frac{N_\xi}{N_\eta}\right).
\]
Integrating and using (\ref{eq:Ns}) we find the expression
\[
A^{(1)}=b\frac{r^3 \sin^2\theta}{r-2m}+const,
\]
which restricted onto $\supo$ reads
\begin{equation}
A^{(1)}|_{\supo}=b\frac{\sigo^3 \asig^3 \sin^2\vartheta}{\asig \sigo-2m}+
const.
\label{eq:cond_static_supo}
\end{equation}
Moreover, the fact that the perturbation $A^{(1)}$ is time-independent
implies that its knowledge on the matching hypersurface $\supo$
implies its knowledge on the whole region swept by the
range of variation of $r_{0} (\lambda)$.
We therefore conclude that 
any interior perturbation which takes the form (\ref{eq:cond_static_supo})
on the boundary 
defines a static perturbation, at least on the range of variation of 
$r_{0} (\lambda)$.

Recalling that both $\sigo$ and $\dersigo$ are constants,
in view of the expressions (\ref{ExactFLRW}) and
(\ref{eq:cond_static_supo}) we have proven:
\begin{theorem}
The most general stationary and axially symmetric first order vacuum
perturbation of a Schwarzschild metric, matching to an exact FLRW
geometry across any linearly perturbed non-null surface, must be static
on the range of variation of $r_{0} (\lambda)$.
\end{theorem}

\section*{Acknowledgements}

MM was supported by the projects
FIS2006-05319 of the Spanish CICyT,
SA010CO5 of the Junta de Castilla y Le\'on and
P06-FQM-01951 of the Junta de Andaluc\'{\i}a.
FM thanks the Dep. Fundamental Physics (Univ. Salamanca) and IMPA (Rio de Janeiro) for hospitality, 
FCT (Portugal) for grant SFRH/BPD/12137/2003 and CMAT, Univ. Minho, for support.
RV thanks the Dep. Fundamental Physics (Univ. Salamanca) for their kind hospitality,
is funded by the Basque Government Ref. BFI05.335,
and acknowledges support from projects FIS2004-01626 from the Spanish MEC
and IT-221-07 from the Basque Government.

\appendix
\section{Perturbed $(-)$ region first and second fundamental forms}
\label{appendixA}
Using the definitions of subsections \ref{intpert}, \ref{intback} and
\ref{intscalars} in expressions (\ref{pertfirst})-(\ref{pertsecond})
we find that the first order perturbation of the induced metric on
$\supo^{-}$ has the following components
\begin{eqnarray}
\ffffpert{}^{-}_{\lambda\lambda}
&=& -2 \left [ \frac{m}{r_0^2} Z^1 \dot t_0^2 + \left ( 1- \frac{2m}{r_0}
\right ) \left ( U^{(1)}  \dot t_0 + \frac{\partial Z^0}{\partial \lambda}\right ) \dot t_0
- \frac{r_0}{r_0 - 2 m}  \dot{r}_0 \frac{\partial Z^1}{\partial \lambda}
+ \frac{m Z^1}{ \left ( r_0 - 2 m \right )^2} \dot{r}_0^2 
\right . \nonumber\\
&&\left. \left. 
+ \frac{r_0}{r_0 -2 m} \left ( U^{(1)} - k^{(1)} \right ) \dot{r}_0^2 
\right ] \right|_{\supo},
\label{eq:q1ll_in}\\
\ffffpert{}^{-}_{\lambda\vartheta}
&=& r_0^2 \frac{\partial Z^2}{\partial \lambda}
- \left ( 1 - \frac{2 m}{r_0} \right ) 
\frac{\partial Z^0}{\partial \vartheta} \dot t_0+ 
\frac{r_0}{r_0 - 2 m} \dot{r}_0 \frac{\partial Z^1}{\partial \vartheta}
, \nonumber\\
\ffffpert{}^{-}_{\lambda\varphi}
&=& 
r_0^2 \sin^2 \vartheta \frac{\partial Z^3}{\partial \lambda}
- \left ( 1 - \frac{2 m}{r_0} \right ) 
\left ( \frac{\partial Z^0}{\partial \varphi} + A^{(1)}|_{\supo} 
\right )\dot t_0+
\frac{r_0}{r_0 - 2 m} \dot{r}_0  \frac{\partial Z^1}{\partial \varphi}
, \nonumber\\
\ffffpert{}^{-}_{\vartheta\vartheta}
&=& \left.2 r_0 Z^1  + 2 r_0^2 \left ( 
\frac{\partial Z^2}{\partial \vartheta} + k^{(1)} 
- U^{(1)}  \right )\right|_{\supo}, \nonumber\\
\ffffpert{}^{-}_{\vartheta\varphi}
&=& r_0^2 \left ( \frac{\partial Z^2}{\partial \varphi}
+ \sin^2 \vartheta \frac{\partial Z^3}{\partial \vartheta} \right )
, \nonumber\\
\ffffpert{}^{-}_{\varphi\varphi}
&=& 2 r_0 Z^1 \sin^2 \vartheta 
+2 r_0^2 \sin \vartheta \left ( \sin \vartheta
\frac{\partial Z^3}{\partial \varphi}
- \uperonepsup|_{\supo} \sin \vartheta  +  Z^2\cos \vartheta \right ), \nonumber
\end{eqnarray}
and that the first order perturbation of the extrinsic curvature has the
following components
\begin{eqnarray}
\sfffpert{}^{-}_{\lambda\lambda}
&=&
\frac{1}{N_{-}} \left (
- \frac{ \left ( r_0 -2m \right )^2}{r_0^2} \dot t_0^3
\frac{\partial U^{(1)}}{\partial r} 
+ \frac{ m \left ( r_0 -2m \right )}{r_0^3} 
\dot t_0^3 \left ( k^{(1)}  - 3 U^{(1)}  \right )
- \frac{2 m \left ( r_0 -2 m \right )}{r_0^3} \dot t_0^2\frac{\partial Z^0}{\partial\lambda} 
\right .  \nonumber\\ & &  \left .  
-\dot t_0\frac{\partial^2 Z^1}{\partial \lambda^2} + \dot t_0^3\frac{m \left ( 2 r_0 - 5 m
\right )}{r_0^4} Z^1 + \dot t_0\dot{r}_0^2 \left ( 3 \frac{\partial U^{(1)}}{\partial r} -  
\frac{\partial k^{(1)}}{\partial r} + \dot t_0^2\frac{2 m  }{N_{-}^2 r_0^2} k^{(1)} \right ) 
\right .  \nonumber\\ & &  \left .  
+ \frac{\dot t_0\ddot{r}_0}{ N_{-}^{2}} \left ( 
\frac{r_0 \dot{r}^2_0}{r_0 - 2 m } \frac{\partial Z^0}{\partial \lambda}
- \frac{r_0 \dot{r}_0}{r_0 - 2 m } \frac{\partial Z^1}{\partial \lambda}\dot t_0
- \dot t_0^3\left ( 1-  \frac{2m}{r_0} \right ) k^{(1)}   \right ) 
+ \dot{r}_0  \frac{\partial^2 Z^0}{\partial \lambda^2} 
\right .  \nonumber\\
&& + \left . \dot t_0 \ddot{r}_0 \frac{ (r_0 - 2m)^2 \dot t_0^2+ r_0^2 \dot{r}_0^2}{(r_0 - 2m)^2 \dot t_0^2-
r_0^2 \dot{r}_0^2} \left (
U^{(1)} + \frac{ m Z^1}{r_0 \left ( r_ 0- 2m \right )} \right ) 
+ \frac{m \dot{r}^2_0}{ (r_0 - 2m )^2\dot t_0^2 - r_0^2 \dot{r}_0^2} 
\times
\right .  
\nonumber\\ & &  \left .  
\left (
\dot t_0 Z^1 \left ( \frac{ 5 m - 6 r_0}{r_0^2}\dot t_0^2 + 3 \dot{r}_0^2
\frac{2 r_0 - 3 m}{ ( r_0 - 2 m )^2} \right ) 
+ \frac{ \dot t_0^2( r_0 -2 m )^2 - 3 r_0^2 \dot{r}_0^2 }{r_0 \left ( r_0 - 2 m \right )} \frac{\partial Z^0}{\partial \lambda}
\right .  \right . \nonumber\\ && \left. \left . \left . 
-  \dot t_0\frac{ ( r_0 -2 m )^2\dot t_0^2 + 3 r_0^2 \dot{r}_0^2 }{r_0 \left ( r_0 - 2 m \right )}
U^{(1)} \right )  
\right  . \right . \nonumber \\
& & 
+  \left . \left . 
\frac{m \dot t_0\dot{r}_0}{ \dot t_0^2(r_0 - 2m )^2 - r_0^2 \dot{r}_0^2} 
\frac{ 5 \dot t_0^2( r_0 -2 m )^2 - 3 r_0^2 \dot{r}_0^2 }{r_0 \left ( r_0 - 2 m \right )} \frac{\partial Z^1}{\partial \lambda}
\right )\right|_{\supo},
\label{eq:k1ll_in} \\
\sfffpert{}^{-}_{\lambda\vartheta}
&=& 
\frac{1}{N_{-}}
\left ( - \dot t_0\frac{\partial^2 Z^1}{\partial
\lambda \partial \vartheta} 
 + \dot t_0\left ( r_0 - 2 m \right) \left ( \frac{\partial Z^2}{\partial \lambda}
- \frac{m}{r_0^3} \frac{\partial Z^0}{\partial \vartheta}\dot t_0 \right )
+ \dot t_0\dot{r}_0 \frac{1}{r_0 - 2 m} \frac{\partial Z^1}{\partial \vartheta} \right . \nonumber\\
&&\left.
\left . + \dot{r}_0 \left ( \frac{\partial^2 Z^0}{\partial \lambda \partial \vartheta}
+ 2 \dot t_0\frac{\partial U^{(1)}}{\partial \theta} -  
\dot t_0\frac{\partial k^{(1)}}{\partial \theta} \right )
- \frac{ \left ( r_0 - 3m \right ) \dot{r}_0^2}{r_0 \left ( r_0 -2 m \right )} 
\frac{\partial Z^0}{\partial \vartheta}
\right )\right|_{\supo},
 \nonumber\\
\sfffpert{}^{-}_{\lambda\varphi}
&=& 
\frac{1}{N_{-}} 
\left ( 
- \dot t_0\frac{\partial^2 Z^1}{\partial \lambda \partial \varphi}
 + \left ( r_0 - 2 m \right) 
\left ( \dot t_0\sin^2 \vartheta \frac{\partial Z^3}{\partial \lambda}
- \frac{m}{r_0^3}\dot t_0^2 \frac{\partial Z^0}{\partial \varphi} \right ) 
- \frac{ \left ( r_0 - 2 m \right )^2}{2 r_0^2} \dot t_0^2  \frac{\partial A^{(1)}}{\partial r}  
\right .  \nonumber\\
&&\left . + \dot{r}_0 \left ( \frac{\partial Z^0}{\partial \lambda \partial \varphi}
+ \frac{1}{r_0 - 2 m}\dot t_0 \frac{\partial Z^1}{\partial \varphi} 
- \frac{ \left ( r_0 - 3m \right ) \dot{r}_0}{r_0 \left ( r_0 -2 m \right )} 
\frac{\partial Z^0}{\partial \varphi}
\right ) - \frac{m \left ( r_0 - 2m \right )}{r_0^3} \dot t_0^2 A^{(1)} 
\right  . \nonumber\\
&&\left.
\left . 
+ \dot{r}_0 \left ( \frac{\dot{r}_0}{2} \frac{\partial A^{(1)}}{\partial r}
- \frac{\left ( r_0 - 3 m \right ) \dot{r}_0 }{r_0 \left ( r_0 - 2 m \right )}  A^{(1)}
\right )
\right )\right|_{\supo}, \nonumber\\
\sfffpert{}^{-}_{\vartheta\vartheta}
&=& 
\frac{1}{N_{-}} 
\left ( 
- \dot t_0\frac{\partial^2 Z^1}{\partial \vartheta^2} 
+ \dot t_0\frac{r_0 - m}{r_0} Z^1
+ \dot t_0\left ( r_0 -2 m \right ) \left ( 
2 \frac{\partial Z^2}{\partial \vartheta}
+ r_0  \frac{\partial k^{(1)}}{\partial r} 
- r_0 \frac{\partial U^{(1)}}{\partial r} 
+ \right . \right . \nonumber\\
&&\left . \left .  \frac{}{}  k^{(1)} - U^{(1)}   \right )
+ \dot{r}_0 \left ( \frac{\partial^2 Z^0}{\partial \vartheta^2} \right )
+ \frac{\dot t_0r_0 \dot{r}_0^2}{  N_{-}^{2}} \left (  
- \frac{\partial Z^0}{\partial \lambda}\frac{1}{\dot t_0} 
+ k^{(1)} - 2 U^{(1)}\right ) \right . \nonumber\\
&&\left.
\left . 
- \dot t_0\frac{ 2 m \dot{r}_0^2 }{N_{-}^2 (r_0 - 2 m)} Z^1
+ \frac{\dot t_0 r_0 \dot{r}_0}{  N_{-}^{2}}   \frac{\partial Z^1}{\partial \lambda} 
\right )\right|_{\supo}, \nonumber\\
\sfffpert{}^{-}_{\vartheta\varphi}
&=& 
\frac{1}{N_{-}} 
\left ( 
- \dot t_0\frac{\partial^2 Z^1}{\partial \vartheta \partial \varphi}
+ \dot t_0\frac{\cos \vartheta}{\sin \vartheta} 
\frac{\partial Z^1}{\partial \varphi} 
+ \dot t_0\left (r_0 - 2 m \right )  \left ( \frac{\partial Z^2}{\partial \varphi}
+ \sin^2 \vartheta \frac{\partial Z^3}{\partial \vartheta} \right ) \right .  \nonumber\\
&&\left.
\left . + \dot{r}_0 \left ( \frac{\partial Z^0}{\partial \varphi \partial \vartheta} 
+ \frac{1}{2} \frac{\partial A^{(1)}}{\partial \theta}
- \frac{\cos \vartheta}{\sin \vartheta} \left ( \frac{\partial Z^0}{\partial \varphi}
+ A^{(1)} \right ) 
\right )  \right ) \right|_{\supo}, \nonumber\\
\sfffpert{}^{-}_{\varphi\varphi}
&=&
\frac{1}{N_{-}} \dot t_0
\left ( 
- \frac{\partial^2 Z^1}{\partial \varphi^2} 
- \sin \vartheta \cos \vartheta \frac{\partial Z^1}{\partial \vartheta}
+ \frac{\left ( r_0 - m \right ) \sin^2 \vartheta}{r_0} Z^1 
- \frac{2 m \dot{r}_0^2 \sin^2 \vartheta Z^1}{N_{-}^2 (r_0  -2 m)} 
\right . \nonumber\\
&&\left. 
+ \left ( r_0 -2 m \right ) \left ( 
2 \sin^2 \vartheta \frac{\partial Z^3}{\partial \varphi} 
+ 2 \cos \vartheta \sin \vartheta Z^2 - \sin^2 \vartheta \left ( 
r_0  \frac{\partial U^{(1)}}{\partial r}  
+ U^{(1)}  + k^{(1)} \right ) \right ) \right . \nonumber\\
&&\left . 
+ \frac{r_0 \dot{r}_0}{ N_{-}^{2}} \sin^2 \vartheta  \frac{\partial Z^1}{\partial \lambda} 
+ \frac{r_0 \dot{r}_0^2}{ N_{-}^{2}} \sin^2 \vartheta  
\left ( k^{(1)} - 2 U^{(1)} - \frac{1}{\dot t_0}
\frac{\partial Z^0}{\partial \lambda} \right ) 
\right . \nonumber \\
& & \left .
\left. 
+ \frac{\dot{r}_0}{\dot t_0} \left ( \frac{\partial^2 Z^0}{\partial \varphi^2}
+ \cos \vartheta \sin \vartheta \frac{\partial Z^0}{\partial \vartheta}  \right )
\right )\right|_{\supo}.\nonumber
\end{eqnarray}

\section{Perturbed $(+)$--region first and second fundamental forms} 
\label{appendixB}
Using the definitions of subsections \ref{extdecomp}, \ref{extback}
and \ref{extscalars} in expressions
(\ref{pertfirst})-(\ref{pertsecond}) we find that the first order
perturbation of the induced metric on $\supo^{+}$ has the following
components
\begin{eqnarray}
\ffffpert{}^{+}_{\lambda \lambda} &=& \gfpert{}^{+}_{\tau\tau}|_{\supo}   - 2 \asig^2 \frac{\partial  \ttauout}{\partial \lambda}
 - 2 \asig \asigdot \ttauout,
\label{eq:q1ll_out}\\
\ffffpert{}^{+}_{\lambda \vartheta}
&=& \gfpert{}^{+}_{\tau\theta}|_{\supo}
- \asig^2 \frac{\partial \ttauout}{\partial \vartheta}
+ \asig^2 \sigo^2 \left ( \frac{\partial^2 \tone}{\partial \lambda \partial \vartheta} 
- \frac{1}{\sin \vartheta} \frac{\partial^2 \ttwo}{\partial \lambda \partial \varphi} \right ) , \nonumber\\
\ffffpert{}^{+}_{\lambda \varphi}
&=& \gfpert{}^{+}_{\tau\phi}|_{\supo}
- \asig^2 \frac{\partial \ttauout}{\partial \varphi}
+ \asig^2 \sigo^2 \left ( \frac{\partial^2 \tone }{\partial \lambda \partial \varphi} 
+ \sin \vartheta \frac{\partial^2 \ttwo}{\partial \lambda \partial \vartheta} \right ) , \nonumber\\
\ffffpert{}^{+}_{\vartheta \vartheta}
&=& \gfpert{}^{+}_{\theta\theta}|_{\supo}   
+2 \asig \sigo \dersigo \cqout 
+ 2 \asig^2 \sigo^2
\left ( \frac{\asigdot}{\asig} \ttauout + 
\frac{\partial^2 \tone}{\partial \vartheta^2} - \frac{1}{\sin \vartheta} \frac{\partial^2 \ttwo}{\partial \vartheta
\partial \varphi} + \frac{\cos \vartheta}{\sin^2 \vartheta} \frac{\partial \ttwo}{\partial \varphi} \right ),
\nonumber\\
\ffffpert{}^{+}_{\vartheta \varphi}
&=& \gfpert{}^{+}_{\theta\phi}|_{\supo}  \nonumber\\
&&
+ \asig^2 \sigo^2 \sin \vartheta \left ( 
 \frac{\partial^2 \ttwo}{\partial \vartheta^2} 
- \frac{1}{\sin^2 \vartheta} \frac{\partial^2 \ttwo}{\partial \varphi^2}
- \frac{\cos \vartheta}{\sin \vartheta}
\frac{\partial \ttwo}{\partial \vartheta} 
+  \frac{2}{\sin \vartheta} \frac{\partial^2 \tone}{\partial \vartheta \partial \varphi} 
-  \frac{2 \cos \vartheta}{\sin^2 \vartheta} \frac{\partial \tone}{\partial \varphi} \right ), \nonumber\\
\ffffpert{}^{+}_{\varphi \varphi}
&=& \gfpert{}^{+}_{\phi\phi}|_{\supo}  
+2\asig\sigo \dersigo \cqout \sin^2 \vartheta
+ 2\asig\sigo^2 \asigdot  \ttauout\sin^2 \vartheta +  \nonumber\\
&&+ 2 \asig^2 \sigo^2
\left ( \frac{\partial^2 \tone}{\partial \varphi^2} 
+ \cos \vartheta \sin \vartheta 
\frac{\partial \tone}{\partial \vartheta} 
+\sin \vartheta \frac{\partial^2 \ttwo}{\partial \vartheta \partial \varphi} 
- \cos \vartheta \frac{\partial \ttwo}{\partial \varphi}  \right ),\nonumber
\end{eqnarray}
and that the first order perturbation of the extrinsic curvature has the
following components
\begin{eqnarray}
\sfffpert{}^+_{\lambda\lambda}
&=&
\left. 
- \frac{1}{\asig} \frac{\partial  \gfpert{}^+_{\tau\rr}}{\partial \tau}
+ \frac{\asigdot}{\asig^2} \left . \gfpert{}^+_{\tau \rr} \right . + 
\frac{1}{2 \asig} \left . \frac{\partial \gfpert{}^+_{\tau\tau}}{\partial \rr} \right . 
- \frac{\partial^2 \cqout}{\partial \lambda^2} 
\right . \nonumber \\ & & \left . 
+ \frac{\asigdot}{\asig} \frac{\partial \cqout}{\partial \lambda} 
+ \left ( \frac{\asigdotdot}{\asig}  - \frac{\asigdot^2}{\asig^2} \right ) \cqout \right|_{\supo},
\label{eq:k1ll_out}\\
\sfffpert{}^+_{\lambda\vartheta}
&=& 
\frac{1}{2 \asig} \frac{\gfpert{}^+_{\tau\theta}}{\partial \rr} -
\frac{1}{2 \asig} \frac{\gfpert{}^+_{\tau\rr}}{\partial \theta} 
+\frac{\asigdot}{\asig^2} \left . \gfpert{}^+_{\rr \theta} \right . -
\frac{1}{2 \asig} \left . \frac{\partial \gfpert{}^+_{\rr\theta}}{\partial \tau} \right . 
- \frac{\partial^2 \cqout}{\partial \lambda \partial \vartheta} + \frac{\asigdot}{\asig} \frac{\partial
 \cqout}{\partial \vartheta}  \nonumber\\
&&\left.+ \asig\sigo \dersigo \left ( 
\frac{\partial^2 \tone}{\partial \lambda \partial \vartheta}
- \frac{1}{\sin \vartheta} \frac{\partial^2 \ttwo}{\partial \lambda \partial \varphi}
\right )\right|_{\supo}, \nonumber\\
\sfffpert{}^+_{\lambda\varphi}
&=& 
\frac{1}{2 \asig} \frac{\gfpert{}^+_{\tau\phi}}{\partial \rr} -
\frac{1}{2 \asig} \frac{\gfpert{}^+_{\tau\rr}}{\partial \phi} 
+ \frac{\asigdot}{\asig^2} \left . \gfpert{}^+_{\rr \phi} \right . -
\frac{1}{2 \asig} \left . \frac{\partial \gfpert{}^+_{\rr\phi}}{\partial \tau} \right . 
- \frac{\partial^2 \cqout}{\partial \lambda \partial \varphi} + \frac{\asigdot}{\asig} \frac{\partial
 \cqout}{\partial \varphi} \nonumber\\
& & \left.+\asig\sigo \dersigo \left ( 
\frac{\partial^2 \tone}{\partial \lambda \partial \varphi}
+\sin \vartheta \frac{\partial^2 \ttwo}{\partial \lambda \partial \vartheta} \right )\right|_{\supo}, \nonumber\\ 
\sfffpert{}^+_{\vartheta\vartheta}
& = &
- \frac{\sigo \dersigo}{2 \asig} \left . \gfpert{}^+_{\rr\rr} \right  .
+ \frac{\asigdot \sigo^2}{\asig^2} \left . \gfpert{}^+_{\tau\rr} \right .  
+ \frac{1}{2 \asig} \left . \frac{\partial \gfpert{}^+_{\theta\theta}}{\partial \rr} \right .
- \frac{1}{\asig} \left . \frac{\partial \gfpert{}^+_{\rr\theta}}{\partial \theta} \right .
\nonumber \\ 
& & 
+ \sigo \dersigo\asig
\left ( \frac{\asigdot}{\asig} \ttauout + 2 \frac{\partial^2 \tone}{\partial \vartheta^2}
- \frac{2}{\sin \vartheta} \frac{\partial^2 \ttwo}{\partial \vartheta \partial \varphi}
+ \frac{2 \cos \vartheta}{\sin^2 \vartheta} \frac{\partial \ttwo}{\partial \varphi} 
\right )
\nonumber\\ & &\left.
- \frac{\partial^2 \cqout}{\partial \vartheta^2} 
+ \frac{\asigdot \sigo^2}{\asig} \frac{\partial \cqout}{\partial \lambda    } 
+ \sigo^2 \cqout \left ( - \curv + \frac{{\dersigo}^2}{\sigo^2} 
- \frac{\asigdot^2}{\asig^2} \right ) 
\right|_{\supo}, \nonumber \\
\sfffpert{}^+_{\vartheta\varphi}
& = & 
\frac{1}{2 \asig} 
\left ( \frac{\partial \gfpert{}^+_{\theta\phi}}{\partial \rr}
- \frac{\partial \gfpert{}^+_{\rr\theta}}{\partial \phi}
- \frac{\partial \gfpert{}^+_{\rr\phi}}{\partial \theta} \right )
+\frac{\cos \vartheta}{\asig \sin \vartheta} \gfpert{}^+_{\rr \phi}
-\frac{\partial^2 \cqout}{\partial \vartheta \partial  \varphi} 
+\frac{\cos \vartheta}{\sin \vartheta} \frac{\partial \cqout}{\partial \varphi} \nonumber\\ & &
\left.
+ \sigo \dersigo\asig\left (
2 \frac{\partial^2 \tone}{\partial \vartheta \partial \varphi} -
2 \frac{\cos \vartheta}{\sin \vartheta} \frac{\partial \tone}{\partial \varphi} 
+ \sin \vartheta \frac{\partial^2 \ttwo}{\partial \vartheta^2 } 
- \cos \vartheta \frac{\partial \ttwo}{\partial \vartheta}
- \frac{1}{\sin \vartheta}
\frac{\partial^2 \ttwo}{\partial \varphi^2} \right ) \right|_{\supo}, \nonumber\\
\sfffpert{}^+_{\varphi\varphi} 
& = &
- \frac{\sin^2 \vartheta \sigo \dersigo}{2 \asig} \left . \gfpert{}^+_{\rr\rr} \right  .
+ \frac{\asigdot \sin^2 \vartheta \sigo^2}{\asig^2} \left . \gfpert{}^+_{\tau\rr} \right .  
+ \frac{1}{2 \asig} \left . \frac{\partial \gfpert{}^+_{\phi\phi}}{\partial \rr} \right .
- \frac{1}{\asig} \left . \frac{\partial \gfpert{}^+_{\rr\phi}}{\partial \phi} \right .
\nonumber\\ & & 
- \frac{\cos \vartheta \sin \vartheta}{\asig} \gfpert{}^+_{\rr\theta}
- \frac{\partial^2 \cqout}{\partial \varphi^2} 
- \frac{\cos \vartheta}{\sin \vartheta} \frac{\partial \cqout}{\partial \vartheta} 
+ \sin^2 \vartheta \frac{\asigdot \sigo^2}{\asig} \frac{\partial \cqout}{\partial \lambda    } 
\nonumber \\
& &
+ \sin^2 \vartheta  \sigo^2 \cqout \left ( - \curv + \frac{{\dersigo}^2}{\sigo^2} -
\frac{\asigdot^2}{\asig^2} \right ) 
+ \sin^2 \vartheta \sigo \dersigo\asig\left ( 
\frac{\asigdot}{\asig} \ttauout 
\right . \nonumber \\
& & \left .  \left . 
+ \frac{2}{\sin^2 \vartheta} \frac{\partial^2 \tone}{\partial \varphi^2}
+ \frac{2 \cos \vartheta}{\sin \vartheta} \frac{\partial \tone}{\partial \vartheta}
+ \frac{2}{\sin\vartheta} \left ( \frac{\partial^2 \ttwo}{\partial \vartheta \partial \varphi}
- \frac{\cos \vartheta}{\sin \vartheta} \frac{\partial \ttwo}{\partial \varphi} \right ) \right )
\right|_{\supo}. \nonumber 
\end{eqnarray}

\section{Linearised matching in Poisson gauge for $\curv=0$}
\label{sec:poisson_gauge}
Here, as an example, we shall write the linearised matching conditions for
the particular case of a flat $\curv=0$ FLRW region in the Poisson
spacetime gauge in FLRW, for which $\wzero=\cchi=\cy_a=0$.

The odd part equations are given by
\begin{eqnarray}  
\label{eq:ttwo_a}
&&\ttwo + \sigo^{-2}[\uytwoP] \eqq  \tintwo, \\
\label{eq:W2_a}
&&\wwtwoP \eqq -  \sigma \cgsup \asig^{-1},\\
\label{eq:dW2_a}
&&\dwwtwoP \eqq -\sigma \cgsup \frac{3m}{\sigo^2 \asig^2}-\sigma\dercgsupr,\\
\label{eq:Q2_a}
&&\qtwo -[\duytwoP] + 2\sigo^{-1}[\uytwoP]
\eqq -\sigma\cgsup \asig^{-2}\asigdot \sigo.
\end{eqnarray}
The even part equations read
\begin{eqnarray}
\label{eq:tone_a}
&&\tone + \sigo^{-2}[\uyoneP] \eqq \tinone + \cpsup,\\
\label{eq:ttau_a}
&&\ttauout  -[\wwoneP] \eqq \sigo^3 \asig \dercpsupr +\ttauin, \\
\label{eq:Q1_a}
&&\qone +\ychi  -[\duyoneP]\nonumber\\
&&~~~~
+2 \sigo^{-1}[\uyoneP]+\diffqq \eqq -\sigo^2 \asig \dercpsupr,\\
\label{eq:Psi_a} 
&& \Psi
+\frac{1}{\asig}\ddl\left[\asig\left.
\left( \wwoneP\right)\right|_{\supo}\right]
\eqq  \nonumber\\
&&~~~~ 
\frac{\asig\sigo + 2 m}{\asig \sigo - 2m}\uperonepsup
-3m \dercpsupr
-2m \der2cpsupr
- \frac{2 m }{ \asig \sigo - 2 m }\opthp,\\
\label{eq:dW1_a}
&&\dwwoneP \nonumber\\
&&~~~~ \eqq
\frac{2\asigdot\sigo^2}{\asig\sigo-2m}(2\uperonepsup-\opthp)
-3\asigdot \sigo^2\dercpsupr-\asigdot\asig\sigo^3\der2cpsupr, \\
\label{eq:Psiprime_a}
&&\derxipsi+\frac{1}{\asig}
\ddl\left(\asig\left[\wchi+\ddl\left(\diffqq\right)\right]|_{\supo}\right)
\eqq
-\frac{3m}{\asig^2\sigo^3}\cqin 
+ \frac{\asig\sigo-6m}{\asig\sigo-2m}
\asig \deruperonepsupr
\nonumber \\
&&~~~~ +\frac{2m}{\asig\sigo-2m}\asig
\frac{\partial}{\partial r}\opthp    
+\frac{2\asig}{(\asig\sigo-2m)^2} (2\uperonepsup-\opthp),
\\
\label{eq:Phiprime_a}
&&\derxiphi
+ \frac{1}{\sigo}\Phi
-\frac{\asigdot}{\asig}\wchi
- \frac{\asigdot}{\asig\sigo}\left(\wwoneP\right) \nonumber\\
&&~~~~
-\frac{1}{\sigo^3}[\ch + \DeltaS\uone]
-\frac{1}{2\sigo^2}[\derxich - {\markkern 4\asig \Pextm Y_1^m}] 
-\ychi\sigo^{-2}
\eqq \nonumber\\
&&~~~~
\frac{1}{\sigo^2}\diffqq
+ \frac{2m}{\asigdot \sigo^3} \ddl\left(\diffqq\right)
-\frac{3m}{\asig^2\sigo^3}\cqin
+\frac{\asig\sigo+2m}{\sigo(\asig\sigo - 2m)}
\uperonepsup
+\asig \deruperonepsupr
\nonumber\\
&&~~~~
-\frac{1}{\sigo} 2m\dercpsupr
+\frac{\asig\sigo-4m}{\sigo(\asig\sigo-2m)}\opthp
+\frac{1}{\sigo}\frac{\cos\theta}{\sin\theta}
\frac{\partial}{\partial\theta}\left(2\cpsup+\asig\sigo\dercpsupr\right),\\
\label{eq:diffqq_a}
&&\diffqq+\frac{\asigdot\sigo}{\asig}\left(\wwoneP\right)\nonumber\\
&&~~~~
+\frac{1}{2\sigo}(\ch-{\markkern 2\Pqm Y_1^m})
+\ychi
-\sigo\Phi
\eqq\sigo\left(-2m\dercpsupr
-\frac{\cos\theta}{\sin\theta}\frac{\partial \cpsup}{\partial \theta}-\uperonepsup\right).
\end{eqnarray}
\section{Identification with Mukohyama's perturbation variables}
\label{variable-identif}
The variables $N, {\cal K}$ and ${\bar {\cal K}}$ here (see subsection \ref{sec:scalar-match})
have been chosen so that they correspond to those of
\cite{Mukohyama} and \cite{MMV1}, while the function $r$
there corresponds to the Schwarzschild radius here, and takes the value
$\asig\sigo$ on the FLRW side of $\supo$.


In the FLRW region, the $S^2$ scalar variables
used here 
are related to the variables in \cite{Mukohyama,MMV1}
in the following way:
\begin{eqnarray*}
  & & \sum^\infty_{l=0} h^+_{00} Y =-2a^2\Psi\nonumber\\
  & & \sum^\infty_{l=0} h^+_{01} Y = a^2 (\wchi +\wzero)\nonumber\\
  & & \sum^\infty_{l=0} h^+_{11} Y = a^2(-2\Phi+\cchi)\nonumber\\
  & & \sum^\infty_{l=0} h^+_{(Y)} Y = a^2[-2\Phi\sig^2
         +\frac{1}{3}\sig\dersig \derxichi +\frac{1}{6} \DeltaS\cchi
         -\frac{1}{3}\sig^2\derxixichi+2\sig\dersig\ychi
         +\DeltaS(\uone+\yone)+\ch]\nonumber\\
  & & \sum^\infty_{l=1} h^+_{(L)_0} Y =
         \left. a^2(\wzero+\wone)\right|_{l\geq 1}\\
  & & \sum^\infty_{l=1} h^+_{(T)_0} Y  =
         \left. a^2 \wtwo\right|_{l\geq 1}\nonumber\\
  & & \sum^\infty_{l=1} h^+_{(T)_1} Y = \left. a^2 (\qtwo+\derxiytwo
         -2\frac{\dersig}{\sig}\ytwo)\right|_{l\geq 1}\nonumber\\
  & & \sum^\infty_{l=1} h^+_{(L)_1} Y = \left. a^2 (\derxichi+\qone+\ychi
         +\derxiyone-2\frac{\dersig}{\sig}\yone
         -\frac{\dersig}{\sig}\cchi)\right|_{l\geq 1}\nonumber\\
  & & \sum^\infty_{l=2} h^+_{(LT)} Y  =
         \left. a^2(\utwo+\ytwo)\right|_{l\geq 2}\nonumber\\
  & & \sum^\infty_{l=2} h^+_{(LL)} Y  =
         \left. a^2(\uone+\yone+\frac{1}{2}\cchi)\right|_{l\geq 2},\nonumber
\end{eqnarray*}
where
$Y$ stands for the corresponding spherical harmonic for any given pair
$(l,m)$, and where the sum over the values of $m$ is to be understood.
Also note that, as explicitly indicated,
the expressions on the right hand side must be
restricted to the corresponding values of $l$.

At either $(\pm)$ side on $\supo$,
the decomposition of
the first fundamental form perturbation tensor $\ffffpert$
in \cite{Mukohyama,MMV1},
is related to the Hodge scalars decomposition used here by
\begin{eqnarray}
  & & \sum^\infty_{l=0} \sigma_{00}   Y =
          \ffffpert{}_{\lambda\lambda}\nonumber\\
  & & \sum^\infty_{l=1} \sigma_{(T)0} Y =
          \left. G^{q}\right|_{l\geq 1}  \nonumber\\
  & & \sum^\infty_{l=1} \sigma_{(L)0} Y=
          \left. F^{q}\right|_{l\geq 1}\nonumber\\
  & & \sum^\infty_{l=2} \sigma_{(LT)} Y=  \left. R^{q}\right|_{l\geq 2} \\
  & & \sum^\infty_{l=0} \sigma_{(Y)}  Y= H^{q} +\DeltaS P^{q} \nonumber\\
  & & \sum^\infty_{l=2} \sigma_{(LL)} Y  =
          \left. P^{q} \right|_{l\geq 2}.\nonumber
\end{eqnarray}
Analogously, the relations regarding
the second fundamental form perturbation
$\sfffpert$ follow from the above replacing $\sigma$ by $\kappa$
on the left hand side, and $q$ by $k$ on the superscripts on the right hand side
quantities.

Finally, the perturbation of the vector $\vec Z$ at either side
is represented in \cite{MMV1} by
\[
\sum^\infty_{l=0} \tilde Q Y    = Q,~~~~~
 \sum^\infty_{l=0} z_\lambda Y  = N^2 T,~~~~~
 \sum^\infty_{l=1}z_{(T)} Y = {\cal T}_1 |_{l\geq 1} ,~~~~~
 \sum^\infty_{l=1}z_{(L)} Y = {\cal T}_2 |_{l\geq 1},
\]
where $\tilde Q$ stands for the $Q$ used in \cite{MMV1}.

\end{document}